\documentclass{aa}
\usepackage{psfig}
\begin{document}

	\thesaurus{08    % A&A Section 6: stars
              (02.13.1;  % MHD,
               06.19.2;  % (Sun:) solar wind,
               08.06.2;  % Stars: formation,
               08.13.2;  % Stars: mass-loss,
               09.10.1)} % ISM:jets and outflows.
	\title{Outflows from magnetic rotators}
	\subtitle{I. Inner structure}
	\author{T. Lery $^{1}$,$^{2}$,
	J. Heyvaerts $^{1}$ , S. Appl $^{1}$,$^{3}$,
	\and 
	C.A. Norman $^{4}$
		}
	\offprints{T. lery}
	\mail{lery@astro.queensu.ca}
	\institute {$^{1}$ Observatoire de Strasbourg
 	   11 rue de l'Universit\'e
	   67000 Strasbourg, France
	\and 
	   $^{2}$ Department of Physics,
	   Queen's University, Kingston,
	   Ontario, K7L 3N6, Canada
	\and
	   $^{3}$ Institut fuer Angewandte Mathematik, 
	   Universitaet Heidelberg, 
	   Im Neuenheimer Feld 293, D-69120 Heidelberg, Germany
	\and 
           $^{4}$ Space Telescope Science Institute
           and Johns Hopkins University
           3700 San Martin Drive, Baltimore, MD 21218, USA
		}

\date{Received December 1997; accepted }

\titlerunning{Outflows from magnetic rotators. I}
\authorrunning{T.Lery et al.}
	\maketitle
	\begin{abstract}

A simplified model for the stationary, axisymmetric structure of 
magnetized winds with a polytropic equation of state is 
presented. The shape of the magnetic surfaces is assumed 
to be known (conical in this paper) within the fast magnetosonic
surface. The model is non-self-similar. 
%************************************************************************
%REFEREE ****************************************************************
Rather than solving the equilibrium perpendicular to the flux surfaces 
everywhere, solutions are found at the Alfv\`en surface where it takes the 
form of the Alfv\`en regularity condition and
at the base of the flow. This constraints 
the Transfield equilibrium in that the Alfv\`en
regularity condition is imposed and the regularity of
the magnetic surfaces at the Alfv\`en critical surface 
is ensured. 
%REFEREE ****************************************************************
%************************************************************************
The model imposes  
criticality conditions at the slow and fast magnetosonic 
critical points using the Bernoulli equation.
These Alfv\'en regularity and criticality 
conditions  are used to evaluate three constants of motion, 
the total energy, angular momentum, and the ratio 
of mass to magnetic flux $\alpha$, as well as the shape of the 
critical surfaces. The rotation rate $\Omega$  and the 
polytropic constant $Q$ as a function of the magnetic 
surfaces, together with the mass-to-magnetic flux ratio 
on the axis $\alpha_0$ entirely specify the model.
Analytic results are given for limiting cases, and parameter
 studies are performed by numerical means. The model 
accepts any boundary conditions. Numerical calculations 
yield the value of the rotation parameter $\omega$.
Rotators can be defined as slow, intermediate or fast 
according to whether $\omega$ is much less or close to 
unity or near its maximum value for fast rotators, 
$(\frac{3}{2})^\frac{3}{2}$. Given the properties of 
astrophysical objects with outflows, the model allows 
their classification in terms of the rotation parameter.
Critical surfaces are nearly spherical for slow rotators, 
but become strongly distorted for rapid rotators. 
The fast point remains at a finite distance for finite 
entropy flows, in contrast to cold flows.It is found that 
for a given mass loss rate, the rotation rate is limited.

\end{abstract}

\section{Introduction}

Outflows are observed in a large class of astrophysical 
objects, from winds emanating from  stars of all spectral 
types, to well collimated jets originating 
from young stars, compact objects and active galactic nuclei. 
Magnetically driven models are among those considered most promising 
for describing the generation of jets.
In this case a large fraction of the stellar or accretion disc's
rotational energy  is converted into 
electromagnetic energy, and subsequently into 
kinetic energy. The influence of the magnetic field, 
and of the rotation on the flow, has to be assessed even 
for other types of objects in which
the magnetic field plays a secondary role 
in the acceleration of the flow. The aim of this 
paper is to analyze the structure of outflows from rotating 
magnetized objects.

Magnetized outflows start close to the central object. 
The mass loss rate is determined by the magnetic field configuration 
in this region and by the boundary conditions. 
After a major acceleration phase, the collimation phase roughly
starts at about the Alfv\'en surface where  
the flow is deflected towards the axis by the so-called hoop stress.
The problem of determining the stationary two-dimensional structure 
of the collimation region of magnetohydrodynamical outflows has not been 
solved. It requires the solution of the equilibrium of forces perpendicular 
and parallel to the magnetic surfaces. 
One can describe the former by using the transfield or 
Grad-Shafranov equation and the later using 
the Bernoulli equation for a polytropic equation of state.
This situation is complicated by the
existence of three critical surfaces which determine three constants 
of motion. Their position is only obtained as part of the global solution.

Various approaches have been applied to address this problem. 
One approach  assumes the shape of the field lines to be given, thus ignoring
the transverse equilibrium (Weber \& Davis (\cite{wd}), 
Mestel (\cite{mestel}), 
Belcher \& Mc Gregor (\cite{belcher}), Hartmann \& Mc Gregor 
(\cite{hart}), 
Pudritz \& Norman (\cite{PN}), Mestel \& Spruit (\cite{ms}), 
Cassinelli (\cite{cassi})). 
The Bernoulli equation is then solved, together with two constants of motion, 
giving the mass to magnetic flux ratio and the total energy, 
which are obtained from the slow and fast magnetosonic critical surfaces. 
This allows the asymptotic speeds of the outflow to be determined, 
but it does not give any information on the collimation. 
A variation of this approach are perturbative 
treatments of a spherically symmetric solution 
used for numerical simulations (Suess \& Nerney (\cite{suess})).
For example, the treatment of an extended version of the Weber-Davies model
in the space surrounding an axisymmetric system
Sakurai(\cite{saku1}),(\cite{saku2}) and  
Uchida \& Shibata (\cite{uchida}) used a fully iterative
numerical method to solve the Bernoulli equation together 
with the transversal force balance. 

Another approach uses the 
%referee **************
assumption
%referee **************
 of self-similarity 
where some specific dependence of the flow variables 
on the independent variables is assumed (Chan \& Henriksen (\cite{chan}),
Blandford \& Payne (\cite{bp}), Pelletier \& Pudritz (\cite{pellpud}), 
Tsinganos \& Sauty (\cite{tsin1}), Sauty (\cite{sauty1}), Sauty \& Tsinganos 
(\cite{sautyt}),  Ouyed \& Pudritz (\cite{ouyedp}), Contopoulos 
\& Lovelace (\cite{conto}),
Henriksen \& Valls-Gabaud (\cite{henrik}), Tsinganos et Trussoni 
(\cite{tsin2}),  
Fiege \& Henriksen (\cite{fiege})). These models account for the 
force balance, but usually are not regular or valid in all space, 
and do not properly account for the fast magnetosonic critical surface. 
The axis of symmetry often appears as a singularity for the 
electrical current (Blandford \& Payne (\cite{bp}), 
Pelletier \& Pudritz (\cite{pellpud})), though Pelletier \& 
Pudritz found a solution where 
the poloidal current did not diverge along the pole and 
at infinity. These solutions correspond to a current free plasma,
where all the necessary poloidal current is concentrated 
along the polar axis. One should also mention the variational 
approach presented by Rosso \& Pelletier (\cite{rosso}), and the 
slender jet approximation developed by Koupelis \& Van Horn (\cite{koupe}) 
and  Koupelis (\cite{koupe2}).

The disk-wind 
connection has also been central in many explanations of
 the origin of outflows. In some the
engine responsible for the emission of the wind 
is a keplerian disk
threaded by a magnetic field that is either generated
in situ or advected-in from larger scales (Blandford \& Payne (\cite{bp}), 
K\"onigl (\cite{ko}), Pelletier \& Pudritz (\cite{pellpud}), 
Wardle \& K\"onigl (\cite{wk}), 
Li (\cite{li}), Ferreira \& Pelletier (\cite{ferr1}),(\cite{ferr2}), 
Ferreira (\cite{ferr3})).
In another type of model, a X-wind is postulated; the interaction
of a protostar's magnetosphere with its surrounding disk
opens some of the magnetospheric
field lines to create the magnetized stellar wind
(Shu et al.(\cite{shul}),
Shu et al. (\cite{shu94}), Najita \& Shu (\cite{naji})). Another method 
consists in studying numerically time dependent evolution of the 
interaction between the source and the flow
(Ouyed  (\cite{ouyed}), Ouyed \& Pudritz (\cite{ouyedp})).

A more general approach is concerned with rigorous theorems on the asymptotic 
structure of magnetized winds (Heyvaerts \& Norman (\cite{HN})), 
where it has been
shown that outflows either become cylindrical at large distances 
from the source or parabolic, depending on whether they carry an electric 
current to infinity or not.

This is intended to be the first of a series of 
papers on the structure of MHD outflows. 
We propose a simplified model based on the assumption 
that the magnetic surfaces possess
a shape in the collimation zone which is known a priori. But unlike the
Weber-Davis type models, the balance of forces perpendicular to the magnetic
surfaces is taken into account 
%************************************************************************
%referee ********************************************************
on the Alfv\'en surface, where it takes the 
form of the Alfv\`en regularity condition, and
at the base of the flow. This constraints 
the transfield equilibrium in that the Alfv\`en
regularity condition is imposed and the regularity of
the magnetic surfaces at the Alfv\`en critical surface 
is ensured. 
%referee *******************************************************
%************************************************************************
Once given two constants of the motion 
describing the rotation and the thermodynamics close to the source
as boundary conditions,
the system of equations allows the determination of 
the three last constants of motion for otherwise general conditions.
In the second paper, we examine collimated outflows using these integrals
of motion to determine the asymptotic structure of the flows.
The third paper will address the question of jet stability with respect to 
magnetic instabilities which are formed in this model. 

This particular paper has a component devoted to our model
and its astrophysical applications,
and a second part which examines this model in context of 
other studies of outflow from magnetic rotators.
The first part formulates the problem (\S 2),
the assumptions and discuss the governing equations
and relevant boundary conditions.
We derive the new set of equations
describing our model and obtain their solution
analytically and numerically 
for slow (\S3), fast (\S 4) and intermediate rotators (\S 5)
and  suggest astrophysical applications.
The second part provides a comparison
between studies and assesses the ability of this model
to allow the classification of rotators (\S 6)
in terms of suitable dimensionless parameters.
It concludes with a summary which highlights the inner 
structure of magnetic rotator outflows (\S 7).

%%%%%%%%%%%%%%%%%%%%%%%%%%%%%%%%%%%%%%%%%%%%%%%%%%%%%%%%%%%%%
\section{Properties of MHD winds}
%%%%%%%%%%%%%%%%%%%%%%%%%%%%%%%%%%%%%%%%%%%%%%%%%%%%%%%%%%%%%

This paper deals with  stationary and axisymmetric
magnetized rotating winds described in the framework of
ideal MHD. 

\paragraph{Basic relations and notations}

Throughout the paper, ($r$,$\theta$,$z$)
denotes cylindrical coordinates around the rotation axis
while $R$ stands for the spherical distance centered on the
wind source. The vectors 
${\bf e}_r$,${\bf e}_{\theta}$,
${\bf e}_z$ are 
the orthogonal unit vectors associated with 
these coordinates.
It is convenient to split the magnetic field and the 
velocity into a poloidal part, which is in the
meridional ($r$,$z$) plane, and a toroidal part.
The former is denoted by a subscript p while the latter is
just the azimuthal component, so that 
\begin{equation}
{\bf B} = {\bf B}_p +B_{\theta} {\bf e}_{\theta}
\label{B}
\end{equation}
The poloidal part
can be expressed in terms of a flux function $a(r,z)$
proportional to the magnetic flux through a circle centered on the
axis passing at point $r$, $z$, as:
\begin{equation}
{\bf B}_p = 
-\frac{1}{r}
\frac{\partial a}{\partial z} {\bf e}_r
+\frac{1}{r}
\frac{\partial a}{\partial r} {\bf e}_z.
\label{Bp}
\end{equation} 

From this it follows that field lines of the 
poloidal field are lines of constant $a$, and so are the 
magnetic surfaces which are the surfaces generated by
rotating them about the polar axis.

The acceleration of gravity, ${\bf g}$ derives from a 
 gravitational potential $G(r,z)$ by 
${\bf g} = -$ {\boldmath $\nabla$} $G$.
The fluid velocity is denoted by ${\bf v}$.
We assume the  density $\rho$ to be related to 
the pressure $P$ by a polytropic equation of state.
This assumption replaces consideration of 
energy balance and is meant 
to represent simply some more complex
heating and cooling processes. Then, we have:
\begin{equation}
 P =  Q \rho^{\gamma}
\label{P}
\end{equation}
where $Q$ is a factor related to the entropy of the flow
and $\gamma$ is the polytropic index.
$Q$ is constant following the fluid motion, but it can vary 
from one flow line to the next.
From the stationary induction equation it 
results that flow lines follow magnetic surfaces. Therefore 
$Q$ is constant on them and is therefore a function of $a$.

A number of equations of stationary axisymmetric ideal MHD 
can be integrated to a set of equations expressing 
the conservation of first integrals following the motion, 
namely the magnetic surface rotation rate $\Omega$, 
the mass flux to magnetic flux ratio $\alpha$ on a 
magnetic surface, the specific energy $E$ and the 
specific angular momentum of escaping matter, $L$.
By the isorotation law and the specific angular momentum
conservation law, the toroidal components of the velocity and
of the magnetic field can be expressed in terms of the 
density and radius as
\begin{equation}
 v_{\theta}  = \frac{L}{r}  + \frac{\rho}{r}
\frac{L - r^2 \Omega}{\mu_o\alpha^2-\rho}
\label{Vtheta}  
\end{equation}
\begin{equation}
 B_{\theta} = \mu_0 \alpha 
\frac{\rho}{r}
\frac{L - r^2 \Omega}{\mu_o\alpha^2-\rho}.
\label{Btheta} 
\end{equation}
These quantities would be singular when 
the denominators vanish, when 
$ \rho = \mu_0 \alpha^2$, unless $r^2$ becomes equal 
to $\frac{L}{\Omega}$  when this happens. It can easily 
be checked that the poloidal velocity becomes in this 
case equal to the local Alfv\`en velocity, 
calculated with the poloidal component of the magnetic 
field, $v_{PA}$.  For this reason $\mu_0 \alpha^2$
 is named the Alfv\`en density
\begin{equation}
 \rho_A \equiv \mu_0 \alpha^2
\label{rhoa}
\end{equation}
and ${L}/{\Omega}$ is the 
square of the so-called cylindrical 
Alfv\'en radius at this Alfv\`en point,
\begin{equation}  
r_{A}^2 \equiv {{L}\over{\Omega}}.
\label{ra2}
\end{equation}
More generally the subscript $A$ will refer 
to values at the 
Alfv\'en point and it can be shown that 
the alfv\`enic Mach number, 
$M_A = {v_P}/{v_{PA}}$, is given by
\begin{equation}
 M_A^2 = {{\rho_A}\over{\rho}}. 
\end{equation}
The Alfv\'enic poloidal speed at the Alfv\'en point can be expressed as
\begin{equation}
v_{PA} \equiv 
\left(
{{\alpha 
|\nabla {\bf a}|}
\over{\rho r}}
\right)_{A}
.
\label{vpa}
\end{equation}

\paragraph{Basic equations}

The projection of the equation of motion on ${\bf B}_P$
yields by integration an equation, the Bernoulli equation,
\begin{equation}
E(a) =
{{v^2}\over{2}} + {{ \gamma}\over{\gamma - 1}} Q \rho^{\gamma - 1}
+ G(r,z) - {{r \Omega B_{\theta}}\over{\mu_0 \alpha}}. 
\label{Bern1}
\end{equation}
This equation can be also expressed,
using equations (\ref{Vtheta}) and (\ref{Btheta}) as 
\begin{eqnarray}
 {{1}\over{2}} {{\alpha |\nabla {\bf a}|^2
}\over{\rho^2 r^2}}&=& E - G
- \frac{\gamma}{\gamma - 1} Q \rho^{\gamma - 1}
+ \rho \Omega^2 {{r_A^2 - r^2}\over{\rho_A - \rho}}
\nonumber \\
& &
- {{1}\over{2}} {{\Omega^2 r_A^4}\over{r^2}}
\left( 1 + {{\rho}\over{r_A^2}}{{r_A^2 - r^2}
\over{\rho_A - \rho}}\right)^2. 
\label{Bern2}
\end{eqnarray}
In the absence
of MHD forces, this first integral would express 
the well known Bernoulli's theorem, i.e. the constancy of 
the sum of the kinetic, enthalpy and gravitational energy 
fluxes. The presence of the magnetic field introduces another
energy flux, the Poynting flux, the fourth term 
in the equation. 

We introduce the following dimensionless quantities
% REFEREE ****************************************************
% REFEREE ***************************************
that depend on the particular fieldline defined by the magnetic flux $a$:
\begin{itemize}
\item The rotation parameter
\begin{equation}
\omega(a) \equiv {{\Omega r_A}\over {v_{PA}}}
\label{omega}
\end{equation}
\item The thermal parameter, closely related to the usual
$\beta$ parameter at the Alfv\'en point
\begin{equation}
\beta(a) \equiv {{2 \gamma}\over{ \left (\gamma-1\right )}}
 {{Q \rho_A^{\gamma-1}} \over {v_{PA}}^2}
\label{beta}
\end{equation}
\item The  wind  energy parameter
\begin{equation}
\epsilon(a) \equiv {{2 E }\over{v_{PA}}^2}
\label{epsilon}
\end{equation}
\item The gravity parameter
\begin{equation}
g(a) \equiv {{2 G M}\over{R_A v_{PA}}^2}.
\label{g1}
\end{equation}
\end{itemize}
% REFEREE ***************************************
% REFEREE ****************************************************
These parameters weight the
importance of associated effects
in terms of the magnetic energy
at the Alfv\'en point.
$\beta$ is a measure of  the importance of
thermal effects in driving the outflow.

%************************************************************************
% REFEREE ****************************************************
The rotation parameter will play a crucial role in this
problem and need to be clearly defined and explained. First
one should take care about the difference  between the definition
given above and a different one used for example by 
Sakurai (\cite{saku1}, \cite{saku2}) or Spruit (\cite{spruit}) 
which is given by 
$\omega = {{\Omega r_A}\over \sqrt{GM/r_A}}$.
The limiting value $\left(\frac{3}{2}\right)^{3/2}$ for $\omega$
corresponds in both cases to the fast rotator limiting case.
It is the minimum energy solution found by Michel (\cite{michel})
for the classical approach of the Weber \& Davies (\cite{wd}) problem
and gives a fast critical point rejected to infinity. 
With the latter definition omega will always be larger than this limiting 
value. The range of variations of $\omega$ is rather different 
with our definition and need some explanations.
For fast magnetic rotators, it can be easily shown in the frame
of the Weber \& Davies (\cite{wd}) study that, neglecting the thermal
effects, i.e. $Q\sim0$, with respect to rotational and magnetic effects,
two solutions exist only for $\omega$ larger or equal to the 
fast rotator limiting value $\left(\frac{3}{2}\right)^{3/2}$. 
If the thermal component of the flow is not neglected,
the rotation parameter $\omega$ still has the same limiting value
but can become smaller than $\left(\frac{3}{2}\right)^{3/2}$ 
and even reach zero in the case of vanishing rotation.
This agrees with the results given by Ferreira (\cite{ferr3}).
Moreover in our model the rotation parameter $\omega$
increases for a given non-vanishing entropy as a function of the
angular velocity and still has the same limiting value
of  $\left(\frac{3}{2}\right)^{3/2}$
corresponding to the very fast rotator limit.
There is actually no contradiction between these results
and the cold wind theory since our limiting value is
compatible with the inequality imposed by $Q=0$.

%$\left(\frac{3}{2}\right)^{3/2}$
%appears in the Weber-Davies regime as an inferior limit
%for a non-vanishing entropy $Q$ but with an infinite angular 
%velocity $\Omega$ and as a consequence the values of $\omega$ 
%superior to  $\left(\frac{3}{2}\right)^{3/2}$ that 
%strictly cold wind theory authorizes 
%can not be physically reached.

% REFEREE ****************************************************
%************************************************************************
The projection of the stationary 
axisymmetric equation of motion perpendicular to the
magnetic surfaces expresses cross-field  
force balance in meridional planes. It is called the 
Grad-Schl\"uter-Shafranov equation or transfield equation
(Okamoto (\cite{oka}), Heyvaerts \& 
Norman (\cite{HN})) and can be written as:
\begin{eqnarray}
& &{{\alpha}\over{\rho r}}\left(
{{\partial}\over{\partial z}} {{\alpha}\over{\rho r}}
{{\partial a}\over{\partial z}}
+ {{\partial}\over{\partial r}} {{\alpha}\over{\rho r}}
{{\partial a}\over{\partial r}}
\right)
- {{1}\over{\mu_0 \rho r}}
\left(
{{\partial}\over{\partial z}} {{1}\over{r}}{{\partial a}
\over{\partial z}}
+ {{\partial}\over{\partial r}} {{1}\over{r}}{{\partial a}
\over{\partial r}}
\right)
\nonumber \\
&=&
E' - {{Q' \rho^{\gamma -1} }\over{\gamma - 1}}
+ {{\alpha'}\over{\alpha}} \ \ {{\mu_0 \alpha^2 \rho}\over{r^2}} \ \ 
{{(L - r^2 \Omega)^2}\over{(\mu_0 \alpha^2 - \rho)^2}} 
\nonumber \\
& &
- {{\rho}\over{r^2}} \ \ 
 {{(L'-r^2 \Omega')(L - r^2 \Omega)}\over{\mu_0 \alpha^2 - \rho}}
- {{L  L'}\over{r^2}}.
\label{Grad}
\end{eqnarray}
It determines the shape of magnetic surfaces.
Primes denote derivatives with respect
to $a$, i.e. $E'= dE/da$.  This is a quasi-linear
partial differential equation for the magnetic 
flux function $a(r,z)$.
Eq. (\ref{Grad}) becomes critical at 
the Alfv\`en surface (Sakurai (\cite{saku1}), Heyvaerts \&
Norman (\cite{HN})) where it looses all its highest order
derivative terms. 
The condition for its solution to be regular at this point
is the so-called the Alfv\'en regularity condition, which
is the particular form assumed by 
the transfield equation at the Alfv\'en point.
It can be written as (see Heyvaerts \& Norman (\cite{HN}) for details)
\begin{eqnarray}
{{\alpha'}\over{\alpha}}  + 2 (1-p) {{r'_A}\over{r_A}}
- 2 (1-p) {{sin \theta_A}\over{r_A \vert \nabla a \vert_A}}
-  {{Q \rho_A^{\gamma -1}}\over{(\gamma - 1)
v_{PA}^{2}}}{{Q'}\over{Q}}&&
\nonumber \\
+{{E'}\over{v_{PA}^{2}}}
+ {{\Omega^2 r_A^2}\over{v_{PA}^{2}}}
\left( {{\alpha'}\over{\alpha}} {{1}\over{(1-p)^2}}
+ 2 {{r'_A}\over{r_A}} {{p}\over{1-p}}
 -{{\Omega'}\over{\Omega}}\right) &= &0.
\label{RegAL}
\end{eqnarray}
In this equation, 
$p$ is the slope of the solution of the Bernoulli equation at the
Alfv\`en point, hereafter called the Alfv\'en slope
\begin{equation}
 p \equiv 1 + \frac{1}{2}\left({{\partial \log \rho}
\over{\partial \log r}}\right)_{A}
\label{pdef}
\end{equation}
that can be expressed as
\begin{equation}
p = 1 - {\frac {\omega }{ \sqrt 
{ \epsilon + g - \omega ^2 - \beta -1 } }} .
\label{Pold}
\end{equation}
MHD flows have two other critical points 
which are brought about by the Bernoulli equation, 
the slow and fast magneto-sonic points which are located 
where the poloidal velocity equals one of the two
magneto-sonic mode speeds.
Any regular solution of the Bernoulli 
equation must pass these points.
Quantities referring to the slow 
or fast magneto-sonic critical 
point will be indicated
by subscripts s or f respectively.
Let the Bernoulli function ${\mathcal B}(r,\rho)$
be the function defined by 
Eq. (\ref{Bern1}), that has a constant value
$E(a)$ following the flow on this magnetic surface. 
The slow and fast critical points 
on a given magnetic surface of flux parameter $a$ 
are located where the differential of  
${\mathcal B}(r,\rho)$ vanishes.  
The vanishing of the differential form of
the Bernoulli equation
at constant $a$ with respect to $\rho$ and $r$
 stands
as the criticality conditions
\begin{equation}
r \frac{\partial  {\mathcal B}}{\partial r }= 0
\label{dBr} 
\end{equation}
\begin{equation}
\rho \frac{\partial {\mathcal B} }{\partial \rho} = 0.
\label{dBrho} 
\end{equation}
So the set of equations is composed of the Bernoulli equation 
and the Alfv\'en regularity condition together with four criticality
conditions, two defined at the slow point and two
at the fast magnetosonic point.

%%%%%%%%%%%%%%%%%%%%%%%%%%%%%%%%%%%%%%%%%%%%%%%%%%%%%%%%%%%%%
\subsection{A model for the structure of rotating MHD winds}
%%%%%%%%%%%%%%%%%%%%%%%%%%%%%%%%%%%%%%%%%%%%%%%%%%%%%%%%%%%%%

The main idea of our simplified model is to
sacrifice exactness for simplicity, while 
retaining a high degree of generality.
Our model does not impose cross-field 
force balance everywhere
but only at a few important places. First at 
the Alfv\'en surface, where it takes the form of 
the Alfv\'en regularity condition. Second at
the basis of the flow, 
which in the case of a low gas to 
magnetic pressure plasma amounts to assuming a 
uniform flux distribution on a 
small sphere surrounding the wind source. Finally 
transfield force balance can be imposed at infinity.
%************************************************************************
% REFEREE% REFEREE ***************************************
This last point will be done in the second paper, 
allowing the problem to be more self-consistent.
% REFEREE% REFEREE ***************************************
%************************************************************************
We solve  the criticality conditions 
at the slow and fast mode critical points. 
Three first integrals of the motion are then 
determined from these conditions, and the Alfv\'en regularity
condition. The boundary conditions determine the other two.

We moreover simplify 
by assuming the shape of poloidal 
field lines to be known up to the fast magnetosonic point.
We postulate,
%************************************************************************
%REFEREE% REFEREE ***************************************
as a first approximation and for analytical convenience, 
%REFEREE% REFEREE ***************************************
%************************************************************************
that  the deviation from conical shape is small enough 
in the region closer to the wind source than the fast critical point
to consider that the three critical points are aligned on conical magnetic
surfaces.
The shape of magnetic surfaces can indeed be regarded as conical
at distances much smaller than the Alfv\'en radius since the 
kinetic energy of the wind is insufficient to distort
the magnetic field. 
In reality the magnetic and kinetic 
energy start to become comparable at the Alfv\'en point
and the flow  shapes the field at the fast point, 
especially if the fast point happens to be distant 
from the Alfv\'en point. 
We do not assume, though, that the magnetic surfaces 
remain conical beyond the fast critical point. 
%************************************************************************
%REFEREE% REFEREE ***************************************
It is of course a first order approximation as can be seen in Sakurai 
(\cite{saku2}) where  realistic magnetic field lines 
are obtained self-consistently by iterative method starting 
from a conical geometry.
However this strong assumption will be relaxed in future works
where more general forms of magnetic surfaces will be considered.
One should consider this approximation as a first step
in the study of the problem.
%REFEREE% REFEREE ***************************************
%************************************************************************

%%%%%%%%%%%%%%%%%%%%%%%%%%%%
\begin{figure}
\psfig{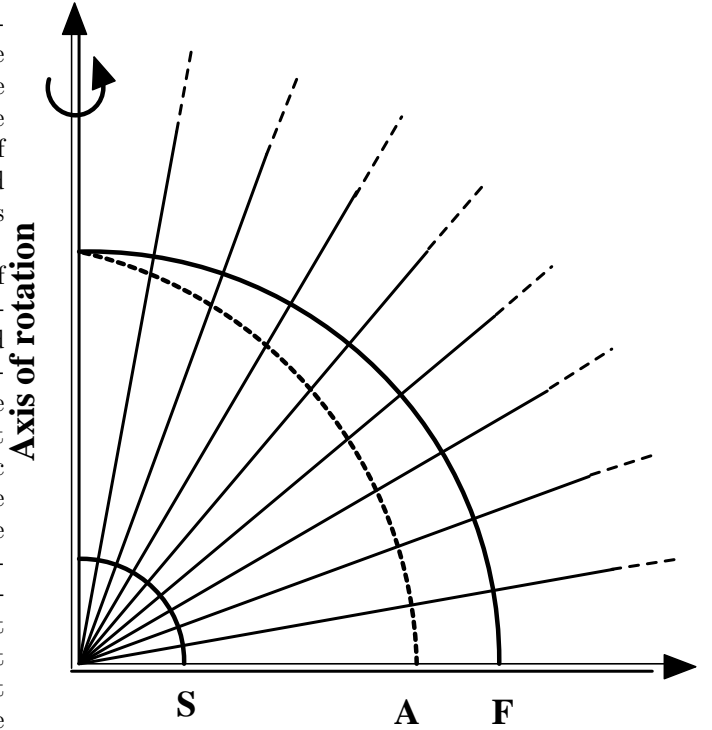}
\caption[ ]{A schematic representation of the structure of the model 
showing the conical shape of magnetic surfaces
in the poloidal plane as light solid lines.
The locus of each of the three critical points 
defines the corresponding
critical surfaces. S, A and F refer to 
the slow magneto-sonic, the alfv\'en and the 
fast magneto-sonic critical surfaces respectively.}
\label{figINNER}
\end{figure}

%%%%%%%%%%%%%%%%%%%%%%%%%%%%%

Consequently a given magnetic surface is represented in this region
by the equation 
$z=r\tan\theta(a)$
and then 
\begin{equation}
r \vert \nabla {\bf a} \vert =
\frac {\cos \theta(a)}{\vert \theta'(a) \vert} .
\end{equation}
For a uniform distribution of the flux,
the relation between the 
angle of a magnetic surface to the equator 
and the flux function $a$
becomes
\begin{equation}
cos^2\theta  = \frac{a}{A} \left( 2 - \frac{a}{A} \right) .
\label{cos}
\end{equation}
Then one can derive all the quantities referring to a magnetic 
surface $a$ defined 
in Eq. (\ref{vpa}),(\ref{omega}),
(\ref{beta}) and (\ref{g1}), only in terms
of the functions $\Omega$, $Q$, $\alpha$, and of the 
distance to the origin and of the density at the Alfv\'en 
point, $R_A$ and $\rho_A$.
\begin{equation}
 v_{PA} = \frac{A}{\mu_0 \alpha R_A^2}
\label{vpa2}
\end{equation}
\begin{equation}
\omega^2 = \frac{\Omega^2 \mu_0^2 \alpha^2 
R_A^6 cos^2\theta}{A^2}
\label{omega2}
\end{equation}
\begin{equation}
\beta = \left(\frac{2 \gamma}{\gamma-1}\right) 
\frac{Q R_A^4 \mu_0^{\gamma+1} \alpha^{2 \gamma}}{A^2}
\label{beta2}
\end{equation}
\begin{equation}
g =  \frac{2 G M \mu_0^2 \alpha^2 R_A^3}{A^2} .
\label{gbis}
\end{equation}

Our set of equations consists of seven equations namely the
Alfv\'en regularity  condition, the four criticality conditions
defined at the slow and fast magneto-sonic points,
and the Bernoulli equation written at the fast and slow critical
points. The model is 
completely determined by the specification of the functions 
$\Omega(a)$, $Q(a)$ and $\alpha$ on the axis ($\alpha_0$). 
The latter variable $\alpha_0$ is directly related and almost 
proportional to the total mass loss rate, and $Q$ and 
$\Omega$ are given by the boundary conditions.

The seven variables which appear in the system of equations  are
$R_{s,A,f}$, $\rho_{s,A,f}$ and $E$. The positions and densities at the 
two critical magnetosonic points $R_s$,  $R_f$, $\rho_s $, and 
$\rho_f$ are given by the four criticality equations. 
The energy $E$ can be eliminated between the two
Bernoulli equations at the fast and slow point.
The resulting equation and the
Alfv\'en regularity equation determine together
$\rho_A$  and $R_A$. The energy $E$ can be calculated
in terms of quantities
defined at the critical points thanks to
the remaining Bernoulli equation
which takes the following form
\begin{eqnarray}
E & = & \frac{1}{2} \frac{\rho_A A^2}{\mu_0 \rho_c R_c^4}
-\frac{GM}{R_c} +\frac{\gamma}{\gamma -1}Q \rho_c^{\gamma-1}
\nonumber \\
 && +\frac{1}{2}\Omega^2 \cos^2\theta\frac{R_A^4}{R_c^2}
\left(1+2\frac{\rho_c}{\rho_A-\rho_c}
\left(1-\frac{R_c^2}{R_A^2}\right)
+\frac{\rho_c^2}{(\rho_A-\rho_c)^2}
\left(1-\frac{R_c^2}{R_A^2}\right)^2
\right)
\label{Ebern}
\end{eqnarray}
where the $R_c$ and $\rho_c$ are the positions and densities
given at the fast or slow critical points.
The positions and densities at the Alfv\'en point are equivalent 
to $L$  and  $\alpha$ via Eq.(\ref{rhoa}) and Eq.(\ref{ra2})
and can be associated to the last two equations of the system.
Thus the three first integrals $E$, $\alpha$ and $L$ are obtained.

%%%%%%%%%%%%%%%%%%%%%%%%%%%%%%%%%%%%%%%%%%%%%%%%%%%%%%%%%%%%%
\subsection{Analytical method}
%%%%%%%%%%%%%%%%%%%%%%%%%%%%%%%%%%%%%%%%%%%%%%%%%%%%%%%%%%%%%

\subsubsection{General procedure}

We can obtain in important limit cases analytical 
expressions for the dimensionless specific energy, 
$\epsilon$ and for the gravity parameter $g$ 
as functions of the parameters $\Omega$, $Q$ and 
$\alpha_0$. We deduce the 
expressions for the positions and densities of 
critical points.
As $\epsilon$ and $g$ have definite values on a given field line,
the expression of these quantities 
as obtained from the slow point criticality
relations and from the fast point criticality relations must 
be the same. 
These constants are functions of
$R_s$, $R_f$, $\rho_s$, $\rho_f$ and,
 of course, of $L$ and $\alpha$ on
magnetic surface $a$.
We use the following dimensionless variables
\begin{equation}
x_s=\frac{r_s}{r_A} \quad , \quad  x_f=\frac{r_f}{r_A}
\label{Xs}
\end{equation}
\begin{equation}
y_s=\frac{\rho_s}{\rho_A} \quad  , \quad  y_f=\frac{\rho_f}{\rho_A} .
\label{Ys}
\end{equation}
In the set of equations given in the previous part
and defining the model, one can calculate the
Bernoulli equation  at the critical points. First 
 at the slow point. From Eq.(\ref{Bern1}) this gives: 
\begin{eqnarray}
{\rm \epsilon}&=&
\beta \, y_s^{\gamma-1}
+\omega^2
\left (2-3 \, x_s^2\right )
\nonumber \\
& &
-{
\frac {3}
{\left (y_s {x_s}^2\right)^2}
}
-\omega^2 
{
\frac 
{ \left (3 x_s^2+1\right )
\left (1 - x_s^2 \right )}
{ x_s^2 \left (1- y_s \right )^2}
} .
\label{epsis}
\end{eqnarray}
Then at the 
fast point (in this case index $s$ is replaced
by $f$). 
\begin{eqnarray}
{\rm \epsilon}&=&
\beta \, y_f^{\gamma-1}
+\omega^2
\left (2-3 \, x_f^2\right )
\nonumber \\
& &
-{\frac {3}{\left (y_f {x_f}^2\right)^2}}
-\omega^2 {\frac { \left (3 x_f^2+1\right )
\left (1-x_f^2\right )}
{ x_f^2\left (1-y_f\right )^2}} .
\label{epsif}
\end{eqnarray}
The positions and densities at the critical points
are obtained from
the vanishing of the differential 
of the Bernoulli equation 
(cf. Eq.\ref{dBr} and Eq.\ref{dBrho}).
This gives the following four equations
\begin{equation}
\frac{\rm g}{2}=
\frac {2}{y_s^2 x_s^3}
+\omega^{2}x_s^3
+\omega^{2}
{\frac 
{\left (1+x_s^2\right )
\left (1-x_s^2\right )}
{x_s {\left (  1-y_s\right )^2}}
}
\label{gs}
\end{equation}
\begin{equation}
\frac{\rm g}{2}=
\frac {2}{y_f^2 x_f^3}
+\omega^{2}x_f^3
+\omega^{2}
{\frac 
{\left (1+x_f^2\right )
\left (1-x_f^2\right )}
{x_f {\left (  1-y_f\right )^2}}
}
\label{gf}
\end{equation}
\begin{eqnarray}
&&\omega^2 x_s^2 \left (1-x_s^2\right )^2
\nonumber \\
&=&\left (
{1-\frac {\beta}{2}}
\left (\gamma-1\right )
x_s^4 y_s^{\gamma+1}\right )
\left (\frac{1-y_s}{ys}\right )^3 .
\label{CRIT1}
\end{eqnarray}
\begin{eqnarray}
&&\omega^2 x_f^2 \left (1-x_f^2\right )^2
\nonumber \\
&=&\left (
{1-\frac {\beta}{2}}
\left (\gamma-1\right )
x_f^4 y_f^{\gamma+1}\right )
\left (\frac{1-y_f}{y_f}\right )^3 .
\label{CRIT2}
\end{eqnarray}
For the isothermal case, the equations 
(\ref{gs}),(\ref{gf}) for $g$
remains the same and only the first term
on the right hand side of the equations
(\ref{epsis}),(\ref{epsif})
for $\epsilon$ are different.
$\beta \, y_s^{\gamma-1}$ and
$\beta \, y_f^{\gamma-1}$ become respectively
$\beta \ln y_s$ and
$\beta \ln y_f$.
And the last two criticality equations are given by
\begin{eqnarray}
&&\omega^2 x_s^2 \left (1-x_s^2\right )^2
\nonumber \\
&=&\left (
{1-\frac {\beta}{2}}
x_s^4 y_s^2\right )
\left (\frac{1-y_s}{ys}\right )^3 .
\label{CRIT1iso}
\end{eqnarray}
\begin{eqnarray}
&&\omega^2 x_f^2 \left (1-x_f^2\right )^2
\nonumber \\
&=&\left (
{1-\frac {\beta}{2}}
x_f^4 y_f^2\right )
\left (\frac{1-y_f}{y_f}\right )^3 .
\label{CRIT2iso}
\end{eqnarray}

%%%%%%%%%%%%%%%%%%%%%
\subsubsection{A general constraint}
%%%%%%%%%%%%%%%%%%%%%

A condition that every solution
has to fulfill can be derived from
Eq.(\ref{CRIT1}), where
all the terms must be positive. Since
$(1-y_f)$ is positive whereas $(1-y_s)$ is negative 
this implies that
\begin{equation}
{\frac {\beta}{2}} \left (\gamma-1\right ) x_s^4 y_s^{\gamma+1}
\geq 1 
\geq 
{\frac {\beta}{2}} \left (\gamma-1\right ) x_f^4 y_f^{\gamma+1} .
\label{IN1}
\end{equation}
From this we derive the constraint 
\begin{equation}
r_{s}^4 \rho_{s}^{\gamma+1} 
\geq 
r_{f}^4 \rho_{f}^{\gamma+1} 
\label{limitXY}
\end{equation}
(For the isothermal case the constraint is given by
$r_{s}^4 \rho_{s}^2 \geq r_{f}^4 \rho_{f}^2 $).
This inequality can also be written
in terms of physically significant quantities as
\begin{equation}
\left[
\frac{\Omega^2 \rho_s r_s^2}{\rho_A}
\right]^2
\left[
\frac{\gamma}{\gamma-1}Q \rho_s^{\gamma-1}
\right]
\geq 
\left[
\frac{\Omega^2 \rho_f r_f^2}{\rho_A}
\right]^2
\left[
\frac{\gamma}{\gamma-1}Q \rho_f^{\gamma-1}
\right] .
\label{INEQphys}
\end{equation}
that is
\begin{equation}
\left[
S^2 H
\right]_{slow}
\geq 
\left[
S^2 H
\right]_{fast} .
\label{INEQphysEN}
\end{equation}
where $S$ is the Poynting specific energy
and H the enthalpy. 
Conditions are thus found on the energies and on the positions and 
densities of the fast and slow magnetic point.
The product of the square of the Poynting flux energy
by the enthalpy at the slow point has to be larger than 
the one calculated at
the fast point. On the other hand,
given the position and the density
at the slow point, the possible values at the 
fast point are restricted by this inequality.

%%%%%%%%%%%%%%%%%%%%%%%%%%%%%%%%%%%%%%%%%%%%%%%%%%%%%%%%%%%%%
\subsection{Numerical method}
%%%%%%%%%%%%%%%%%%%%%%%%%%%%%%%%%%%%%%%%%%%%%%%%%%%%%%%%%%%%%

Numerical investigations can study all ranges of parameters. 
 The variables selected to describe 
the system remain $r_s$, $r_f$, $r_A$,
and $\rho_s$, $\rho_f$ and $\rho_A$
with two free functions, $Q(a)$, 
$\Omega(a)$ and a free parameter, $\alpha_0$, the
value taken by $\alpha(a)$ on the polar axis.
The different forms of  $Q(a)$ and $\Omega(a)$
that have  been considered are given in Appendix B.
It is convenient to convert the algebraic 
magnetosonic criticality
conditions into differential equations.
The critical 
points can be followed from one 
magnetic surface to the next by differentiating 
these criticality equations with respect to $a$.
\begin{equation}
\frac{d}{da} \biggl(\frac{\partial {\mathcal B}}
{\partial r}\biggr)_{s} = 0
\qquad,\qquad
\frac{d}{da} \biggl(\frac{\partial {\mathcal B}}
{\partial \rho}\biggr)_{s} = 0
\label{NUM1S}
\end{equation}
\begin{equation}
\frac{d}{da} \biggl(\frac{ \partial{\mathcal B}}
{\partial r}\biggr)_{f} = 0
\qquad,\qquad
\frac{d}{da} \biggl(\frac{\partial {\mathcal B}}
{\partial \rho}\biggr)_{f} = 0 .
\label{NUM1F}
\end{equation}
Two more equations close the set, namely
the Alfv\`en regularity condition Eq.(\ref{RegAL})
and an equation which expresses the fact
that the specific energy
has a given value on a  magnetic surface,
so that $E(a)= {\mathcal B} (r_s, \rho_s) =
{\mathcal B} (r_f, \rho_f)$
which can be expressed in differential form as
\begin{equation}
\frac{d}{da} ( {\mathcal B}(r_s, \rho_s) -
{\mathcal B}(r_f, \rho_f)) = 0 .
\label{NUM3}
\end{equation}
 The system then consists of six differential equations for
six variables. In practice, equations (\ref{NUM1F}) 
have been multiplied by a factor
$(1-\frac{\rho_f}{\rho_A})$ to prevent singularities
near the axis where $\rho_f$ is almost equal to
$\rho_A$, since this factor appears in the denominators.
A similar problem arises close to the axis and 
on the axis in the calculation of the Alfv\'enic slope
defined by Eq.(\ref{pdef}).
We use the definitions given by 
Eq.(\ref{varpiP}) and Eq.(\ref{VARPI})
to compute this slope 
given a $\omega$ almost
equal to zero near the axis.

The resulting system of equations is then of the form
 \begin{equation}
{\mathcal A}_{ij} \frac{dy_j}{da} = f_i  \qquad , \qquad i,j=1..6
\label{diffA}
\end{equation}
where ${\mathcal A}$ is its  matrix.
We use standard integrators for stiff systems
of initial value problems of first order ordinary 
differential equations. 
We have chosen to study the problem with
initial conditions on the polar axis for simplicity.
Other ways of numerically solving  the problem
could have been chosen.
For example different boundary conditions
could have been imposed for solutions that 
do not fill all space, arguing that thick disks would 
block some part of the available space.
A pressure balance condition
at the external boundary of the outflow
could have been introduced.
Another possibility would have been to prescribe
densities or positions of the critical surfaces
on the equator. 
The solution on the equator would induce the
solution on the polar axis.
 We have chosen to study
systematically problems with
 initial conditions for numerical 
convenience but all the other possibilities
have been tried and indeed converge to the same 
solution for similar parameters.
This study of other boundary conditions for numerical
integration of the system has allowed us
to check the numerical accuracy of the solutions.

The first difficulty in solving this system
consists in evaluating the initial values of the
unknown functions on the axis. We return to this aspect later.
We integrate this system for different values 
of the adjustable functions $\Omega(a)$, $Q(a)$ and 
of the parameter $\alpha_0$, 
the mass to magnetic flux ratio on the axis,
which will be related a-posteriori to
the total mass loss rate.
The function $Q(a)$, related 
to the entropy, is given by boundary
conditions and depends on the object to be modeled.
The rotation rate $\Omega (a)$ of a magnetic
surface is very similar to that of 
the wind-emitting object at the base of the magnetic surface if 
this object is treated as a point source of wind.
Once the positions and densities are computed at the three critical
points it is possible to deduce all the other relevant
variables such as the first integrals $E$, $\alpha$, $L$ or
$\beta$, $\omega$ and the components of the magnetic field
and of the velocity.

\paragraph{Dimensional quantities}
The input parameters of the model can be selected 
so as to reproduce, at least qualitatively, 
observed situations. 
Once given the mass $M_{\ast}$ of the wind-emitting
object, star or disk, the radius $R_{\ast}$, 
the temperature $T_{\ast}$, the density $n_p$, the 
total mass loss rate $\dot M_{\ast}$, the magnetic field
 $B_{\ast}$,
the factor $Q_{\ast}$ and $\gamma$, the dimensionless
parameters  $\overline \Omega$, $\overline Q$, 
$\overline \alpha_0$ can be deduced. 
The parameter  $\overline \alpha_0$ can 
be a posteriori related to
 the mass loss rate $\dot M_{\ast}$, $R_{\ast}$, 
and the magnetic field $B_{\ast}$.
So we define :
\begin{equation}
Q_{\ast}\equiv \frac
{2 k T_{\ast} n_{p_{\ast}}}
{(m_p n_{p_{\ast}})^{\gamma}}
\label{Qstar}
\end{equation}
\begin{equation}
\alpha_{\ast} \equiv \frac
{\dot M_{\ast}}
{4 \pi R_{\ast}^2 B_{\ast}}
\label{Mstar}
\end{equation}
\begin{equation}
\Omega_{\ast}\equiv  \sqrt{\frac
{GM_{\ast}}{R_{\ast}^3}} . 
\label{Omstar}
\end{equation}
All those quantities are nondimensionalized
to reference values by setting
$\overline Q \equiv Q_{\ast}/Q_{ref}$,
$\overline \alpha_0 \equiv \alpha_{\ast}/\alpha_{ref}$ and
$\overline \Omega \equiv \Omega_{\ast}/\Omega_{ref}$.
Major quantities of reference are given by
\begin{equation}
R_{ref} = 8.7\times10^{10}m
\label{RREF}
\end{equation}
\begin{equation}
\rho_{ref} = 3.4\times10^{-18}kg.m^{-3}
\label{RhoREF}
\end{equation}
\begin{equation}
P_{ref} = 5.2\times10^{-9}Pa .
\label{PREF}
\end{equation}
It has been found convenient to start by simply
considering constant values of
$\Omega(a)$ and $Q(a)$. 
In the following subsection the results of our 
model will be illustrated by considering the 
specific examples of the Sun and of a normal TTauri star. 
We shall discuss in the next subsections
the effects of varying rotation, mass loss rate 
and base temperature. Finally we shall show solutions
for non-constant $\Omega$ and $Q$.

%%%%%%%%%%%%%%%%%%%%%%%%%%%%%%%%%%%%%%%%%%%%%%%%%%%%%%%%%%%%%
\section{The Solar wind 
$\left(\omega \ll 1,\Omega=cste\right)$}
%%%%%%%%%%%%%%%%%%%%%%%%%%%%%%%%%%%%%%%%%%%%%%%%%%%%%%%%%%%%%

Let us consider slow magnetic rotators,
like the Solar wind,  which we define as winds for which 
$\omega \ll 1$. 
For slow rotators 
analytical solutions of the criticality 
equations can be found. 
We want to calculate the three first
integrals which result from the imposition of regularity
conditions. In particular, 
the specific energy is expressed as
\begin{eqnarray}
E(a) &=& {{\alpha^2 A^2}\over {2 \rho ^2 R^4}}
 - {{G M} \over R }+
{\gamma\over{\gamma -1}} Q \rho ^ {\gamma-1}
\nonumber \\
& &
+ {{\Omega ^2 cos^2\theta R_A ^4}\over{2 R^2} }
 \left( 1 + {{(1-\frac{R^2}{R_A^2})^2} \over
{(1-\frac{\rho}{\rho_A})^2}}
 {{\rho (2-\frac{\rho}{\rho_A})} \over {\rho_A}}\right) .
\label{DEFe2}
\end{eqnarray}
When $R_A$ is close to $R_f$ we need to evaluate the 
ratios in this expression in a limit sense. 
Therefore to calculate $E$, we need to 
obtain the slope $p$ at the Alfv\`en point
which is also a quantity of interest in the context 
of the Alfv\'en regularity condition, 
Eq.(\ref{RegAL}).
This slope (see Eq. (\ref{pdef})) 
can be expressed as
\begin{equation}
p \equiv 1 - 1/ \varpi.
\label{varpiP}
\end{equation}
with
\begin{equation}
\varpi = 2 {{{A^2\over{R_A^4 \mu_0\rho_A}}-
\gamma Q \rho_A ^{\gamma-1}} \over
{ {{GM} \over{R_A}} - {{2 A^2}\over{R_A^4 \mu_0\rho_A}}
-\Omega^2 cos^2\theta R_A^2 }} .
\label{VARPI}\end{equation}
The specific energy can be evaluated only in terms
of variables defined at the Alfv\`en point
thanks to this variable $\varpi$.
Then we get
\begin{eqnarray}
E(a) &=& {{A^2}\over {2 \mu_0\rho_A R_A^4}}
 - {{G M} \over R_A }+
{\gamma\over{\gamma -1}} Q \rho_A ^ {\gamma-1}
\nonumber \\
& &
+ \frac{\Omega ^2 cos^2\theta R_A ^2}{2} 
 ( 1 + \varpi ^2) 
\label{DEFea}
\end{eqnarray}
The specific energy is then fully determined
in terms of the input parameters.

Assuming a small value for the thermal parameter $\beta$
and eliminating $\epsilon$ and $g$ between 
equations (\ref{epsis}),(\ref{epsif}),
(\ref{gs}) and (\ref{gf}),
we obtain equations for the positions and densities of 
the slow and fast magneto-sonic critical points
relatively to the positions and densities of the 
Alfv\`en point. The position and density
of the slow point exhibits
 interesting behaviors with respect to the different input 
parameters. They are given in the $\omega \ll 1$ limit by
\begin{equation}
{\it {\frac {r_{s}}{r_{A}}}}
=\left (
{\frac 
{\beta \left(5 - 3 {\it \gamma} \right)}
{2 \left(1+\beta-\omega^2 \right) }}
\right )
^{\frac {\gamma+1}{4 ( \gamma-1) }}
\left({\frac {2}
{\beta \left(  \gamma-1 \right)} }
\right)^{1/4}
\label{rsra}
\end{equation}
\begin{equation}
{\it {\frac {\rho_{s}}{\rho_{A}}}}
=\left (
\frac 
{2 (1+\beta-\omega^2) }
{\beta \left (5-3 {\it \gamma}\right )}
\right )
^\frac {1}{{\it \gamma}-1} .
\label{rhosa}\end{equation}
In those equations the polytropic index $\gamma$ 
has two  critical values, namely $\gamma =1$ , 
the isothermal case, that does not create
any problem and can be studied apart, 
and $\gamma =  5/3$.
The latter value is never considered 
since $\gamma$ must be smaller than 
$3/2$ for accelerated winds (Parker(\cite{parker}) ).
When the thermal parameter $\beta$ 
decreases, the slow point moves away
from the Alfv\'en point, approaching  the source 
while the density at this point increases. The dependence on 
$\omega$ is weak, so the slow point is not much affected
by the rotation.
The fast point parameters present an  opposite behavior. We have
\begin{equation}
\left ({\it {\frac {r_{f}}{r_{A}}}}\right)^{-1}
=1-\frac 
{4 \omega^2 
\left (1-\frac {\beta\left ({\it \gamma}-1\right )}
{2}\right )^2}
{
\left (
2+\omega^2-2
\left (
\frac 
{\beta\left (5-3{\gamma}\right )}
{2 \left (1+\beta-\omega^2\right ) }
\right )
^\frac {5-3{\gamma}}
{4 {\gamma}-4}
\left (\frac 
{\beta\left ({\gamma}-1\right )}
{2}\right )^{3/4}
\right )^3
}
\label{rarf}\end{equation}
\begin{equation}
{\it {\frac {\rho_{f}}{\rho_{A}}}}
=1-\frac 
{4 \omega^2 
\left (1-\frac {\beta\left ({\it \gamma}
-1\right )}{2}\right )}
{
\left (
2+\omega^2-2
\left (
\frac 
{\beta\left (5-3{\gamma}\right )}
{2 \left (1+\beta-\omega^2\right ) }
\right )
^\frac {5-3{\gamma}}
{4 {\gamma}-4}
\left (\frac 
{\beta\left ({\gamma}-1\right )}
{2}\right )^{3/4}
\right )^2} . 
\label{rhofa}
\end{equation}

They depend much more on $\omega$ 
than on $\beta$. As $\omega$ increases, the fast 
point separates more and more from the 
Alfv\`en point. 
When the rotation vanishes, the fast point approaches 
the Alfv\'en point and both densities tend to become equal.
Thanks to those results,
analytical expressions can be derived  
for the dimensionless specific energy, $\epsilon$,
and for the gravity parameter, $g$, namely 
\begin{equation}
\epsilon = 1+\beta+3 \omega^2 - {\it g}
\label{epsiLENT}\end{equation}
\begin{equation}
{\it g}
=2^{5/4}\left (\gamma-1\right )^{3/4}
\left (
{\frac 
{5-3 \gamma}
{2\left (1+\beta-\omega^2 \right )}
}\right )
^{
{\frac 
{5-3 \gamma}
{4\left (\gamma-1\right )}}}
\beta^{{\frac {1}{2\left (\gamma-1\right )}
}} .
\label{gLENT}\end{equation}
While  the relative specific energy $\epsilon$ is almost unity 
and depends weakly on $\beta$ and on $\omega$, the parameter
 $g$ shows a clear relation with $\beta$.
Neglecting the effects of rotation and taking  the cold limit, 
the specific energy becomes
\begin{equation}
E = {\frac {A^2}{2 \mu_0 \rho_A R_A^4}} . 
\label{Ecold}\end{equation}
The gravitation parameter tends to vanish as
\begin{equation}
 g  \simeq 
{2}^{5/4}\left (\gamma-1\right )^{3/4}
\left ({\frac {5-3\,\gamma
}{2}}\right )^{{\frac {5-3\,\gamma}
{4\left (\gamma-1
\right )}}}\beta^{{\frac {1}{2\left 
(\gamma-1\right )}}} \ll 1 .
\label{gCOLD}\end{equation}
We need to compute the Alfv\'en slope. We use 
Eq.(\ref{epsiLENT}) in  Eq.(\ref{Pold}).
We find a simple expression for the Alfv\'en slope, valid
in this low-rotation limit
\begin{equation}
  p = 1 - \frac{1}{\sqrt{2}} . 
\label{Preduit}\end{equation}
All the previous results are expressed 
in terms of the position and 
the density at the Alfv\'en point. We now can calculate 
them. Using Eq. (\ref{Preduit}) 
and Eq. (\ref{VARPI}),
we find a simple relation between
the position and the density at the Alfv\'en point
\begin{equation}
\rho_A = 
\biggl( 
\frac
{(1+\sqrt2) A^2}
{\gamma Q R_A^4}
\biggr)
^{1/\gamma} .
\label{rhoa2}
\end{equation}
The position of the Alfv\'en point $R_A$ 
can be obtained by integrating the 
Alfv\'en regularity condition, Eq. (\ref{RegAL})
which introduces a constant of integration $C_1$
related to boundary conditions such as 
the total mass loss rate. 
We have
\begin{equation}
R_A = 
({\mu_0 C_1}{\rho_A E})^{\frac{1}{2\sqrt{2}}} .
\label{Ra1}\end{equation}
From  Eq.(\ref{rhoa}), Eq.(\ref{Ecold}) 
and Eq.(\ref{Ra1}) we can express $R_A$
as
\begin{equation}
R_A = \Bigl( \frac{ C_1 \mu_0 A^2}{2} \Bigr)
^{\frac{1}{2(\sqrt{2}+2)}} .
\label{Ra2}
\end{equation}
Thus in the cold limit it is possible to
derive, from the initial conditions, all the relevant quantities,
namely the positions and densities at the critical points, 
and  the three first integrals of the motion,
with the fast point almost coinciding with
the Alfv\'en point in the cold limit.
So, the problem is fully solved.
For slow rotators, we have
found simple expressions for $\epsilon$, $g$, and for
the Alfv\'en radius and density with the introduction
of a constant of integration given by initial conditions.

%%%%%%%%%%%%%%%%%%%%%%%%%%%%%%%%%%%%%%%%%%%%%%%%%%%%%%%%%%%%%
\subsection{Limit of vanishing rotation }
%%%%%%%%%%%%%%%%%%%%%%%%%%%%%%%%%%%%%%%%%%%%%%%%%%%%%%%%%%%%%

When the rotation vanishes the solution for $Q(a)$ 
independent of $a$ and a uniform flux distribution on the source  
must be spherically symmetric. Moreover the positions of 
the fast magnetosonic
and Alfv\'en points merge in the cold limit. 
We calculate the remaining four unknowns
by making use of the slow point criticality equations. 
After some simple calculations, $\rho_s $ and
$R_s$ are obtained as
\begin{equation}
\rho_s = 
\biggl( \frac 
{16 \alpha^2 A^2 \gamma ^3 Q^3}
{G^4 M^4} \biggr)
^{\frac {1}{5-3\gamma}}
\label{rhosnorot}
\end{equation}
\begin{equation}
R_s = 
\biggl( \frac 
{(GM)^{\gamma+1}}
{(2 \alpha^2 A^2)^{\gamma-1}(2 \alpha Q)^2} \biggr)
^{\frac {1}{5-3\gamma}}
\label{Rsnorot}\end{equation}
The Bernoulli equation written at the Alfv\'en point
gives an equation for $R_A$ 
\begin{equation}
R_A^4 (
E - Q\frac{\gamma}{\gamma-1} \rho_A^{\gamma-1}
+\frac{GM}{R_A}
) = \frac {A^2}{2 {\mu_0}^2 \alpha^2} .
\label{RA4}
\end{equation}
where E can be found from the position
and density at the slow point 
\begin{equation} 
E = \frac {\alpha^2 A^2}{2 R_s^4 \rho_s^2}
-\frac{GM}{R_s}
+Q \frac{\gamma}{\gamma-1} \rho_s^{\gamma-1} .
\label{Es}
\end{equation}
When $\beta$ is small compared to unity, we get
\begin{equation}  
R_A \approx \frac {A^2}{2 {\mu_0}^2 \alpha^2 E}
 \approx \frac{R_s^4 \rho_s^2}{\mu_0^2 \alpha^4} .
\label{Ranorot}
\end{equation}
Using Eq.(\ref{Ra2}), we can express in terms of the 
input parameters the constant $C_1$ which appeared in the  
integration of the Alfv\'en 
regularity relation 
\begin{equation} 
C_1=
\frac{2}{\mu_0 A^2}
\biggl(
\frac{R_s^4 \rho_s^2}
{\mu_0^2\alpha^4}
\biggr)^
{2(\sqrt{2}+2)} .
\label{C1}
\end{equation}
Using the expressions of $R_s$ and $\rho_s$
given by Eq.(\ref{Rsnorot}) and 
Eq.(\ref{rhosnorot}),  this can be also expressed as
\begin{equation} 
C_1=
\frac{2}{A^2 \mu_0^{9+4\sqrt{2}} 
\alpha^{8\left(\sqrt{2}+2\right)}}
\biggl(
\frac{\gamma^{\frac{3}{2}} (GM)^{\gamma-1} A^{3-2\gamma}}
{2^{\gamma-1} Q^{\frac{1}{2}} \alpha^{2\gamma-1}}
\biggr)^
{\frac{8(\sqrt{2}+2)}{5-3\gamma}} .
\label{C1def}
\end{equation}

In the case of vanishing rotation,
it has thus been possible to obtain an
analytical solution.
We got simple 
expressions for the positions and 
densities at the critical points and the constant 
of integration in terms of the input parameters 
only, this including the wind mass loss rate, represented 
by parameter $\alpha_0$. This solution is 
important because it gives initial conditions 
on the axis for 
the numerical integration of
the equations of our model of rotating MHD winds.

%%%%%%%%%%%%%%%%%%%%%%%%%%%%%%%%%%%%%%%%%%%%%%%%%%%%%%%%%%%%%
\subsection{Slow cold rotator}
%%%%%%%%%%%%%%%%%%%%%%%%%%%%%%%%%%%%%%%%%%%%%%%%%%%%%%%%%%%%%

Ignoring $\beta$ in Eq. (\ref{rarf}) and 
Eq. (\ref{rhofa})  we calculate 
the solution for the fast point parameters
\begin{equation}
\rho_f = 
\rho_A \biggl(1 - \frac{4 \omega^2}{(2+\omega^2)^2}\biggr)
\sim \rho_A ( 1 - \omega^2)
\label{rhofCOLD}\end{equation}
\begin{equation}
r_f = r_A 
\biggl(1 - {\frac{4 \omega^2}{(2+\omega^2)^3}}\biggr)^{-1}
\sim r_A 
\left( 1 - \frac{\omega^2}{2}\right)^{-1} .
\label{rAcold}\end{equation}
For $\omega \ll 1$, the relation gives rise to finite
values of the quantities. The fast point remains close to the
Alfv\'en point and depends  on $\omega^2$. 

\subsection{A numerical solution for the Solar wind}

\begin{figure}
\psfig{figure=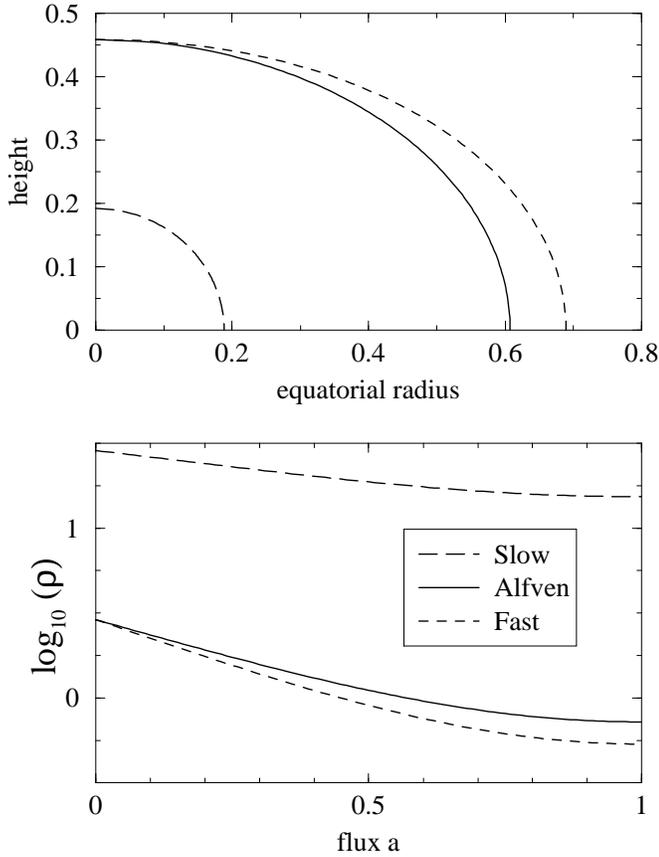,width=\linewidth}
\caption[ ]{
results obtained for 
defined by $Q=2.1$, $\Omega=3.1$ and $\alpha=1.7$ as input 
parameters.  Quantities referring to the Alfv\`en point
are plotted with solid line while dashed lines correspond
to the fast point and long dashed line to the slow point.
}
\label{figSUN}
\end{figure}

Taking into account realistic values for the sun
it has been found that $\omega$ was equal to $0.55$
with $Q=2.1$, $\Omega=3.1$ and $\alpha=1.7$ as input 
parameters. Thus the Sun can be considered as a slow rotator
with respect to our classification, or at least
at the border between slow and intermediate
rotators. The numerical solution
shows clearly that the critical surfaces do not
differ much from spherical shape,
particularly for the slow surface. The fast
magnetosonic point is close to the Alfv\'en point
even near the equator, where the effect of the 
rotation is more important as expected. 
We have also plotted the 
various densities in Fig. \ref{figSUN}.
The fluid near the polar axis is denser than 
near the equatorial part
and one should keep in mind the large values 
of the densities for later comparison with
fast rotators.

%%%%%%%%%%%%%%%%%%%%%%%%%%%%%%%%%%%%
%%%%%%%%%%%%%%%%%%%%%%%%%%%%%%%%

%%%%%%%%%%%%%%%%%%%%%%%%%%%%%%%%%%%%%%%%%%%%%%%%%%%%%%%%%%
\section{Jets from YSOs 
$\left(\left(\frac{3}{2}\right)^\frac{3}{2} - \omega \ll 1
\right)$}
%%%%%%%%%%%%%%%%%%%%%%%%%%%%%%%%%%%%%%%%%%%%%%%%%%%%%%%%%%

%%%%%%%%%%%%%%%%%%%%%%%%%%%%%%%%%%%%%%%%%%%%%%%%%%%%%%%%%%
\subsection{Stellar winds ($\Omega = cste$)}
%%%%%%%%%%%%%%%%%%%%%%%%%%%%%%%%%%%%%%%%%%%%%%%%%%%%%%%%%%

Now, we focus on the  case of fast rigid rotators. 
Fast rotators are defined here as having 
a rotation parameter $\omega$ close
to $\left(\frac{3}{2}\right)^\frac{3}{2}$.
This value
is a maximum for $\omega$ in this conical model and
must be regarded as an asymptotic value
for plasmas rotating at a very large 
physical rotation rate
$\Omega$.
Similarly to the case of the 
slow rotator we can derive analytically in this limit 
the positions and densities at
the critical points relative to 
those at the Alfv\'en point. 
Assuming that $r_s \ll r_A$ and $\rho_s \gg \rho_A$
they are given by
\begin{equation}
{\frac {r_{s}}{r_{A}}} =
\left( \frac{\beta \left(\gamma-1 \right)}{2} \right)^
{\frac{1}{2\left(\gamma-1 \right)}}
\left(
\left(\frac{5-3\gamma}{\gamma-1}\right)
\left({3\omega^{\frac{4}{3}}-2\omega^2}\right)
\right)^
{\frac{\gamma+1}{4\left(\gamma-1\right)}}
\label{RsRa2}
\end{equation}

\begin{equation}
{\frac {\rho_{s}}{\rho_{A}}} =  
\left( \frac{\beta\left(5-3\gamma\right)}
{2}\left(3\omega^{\frac{4}{3}}-2\omega^2\right)
\right)^{-\frac{1}{\gamma-1}} .
\label{rhoas2}
\end{equation}
The slow point parameters depend both on $\omega$ and $\beta$.
When $\beta$ decreases, if the Alfv\`enic point
remains approximately at the same position and density, the slow 
point approaches the source and the corresponding density 
increases. 
At the fast point assuming that 
$r_f \gg r_A$ and $\rho_f \ll \rho_A$
 we have similarly 
\begin{eqnarray}
{\frac {r_{f}}{r_{A}}} &=& \frac 
{1+\frac{3}{\omega^{\frac{4}{3}}}-\frac{4}
{3\omega^{\frac{2}{3}}}}
{\left( \frac{\beta \left(\gamma-1 \right)}{2} \right)^
{\frac{3}{2\left(\gamma-1 \right)}}}
\nonumber \\
& &
\left(
\left(\frac{5-3\gamma}{\gamma-1}\right)^
{\left(\frac{\gamma+1}{4\left(\gamma-1\right)}\right)}
\left(3\omega^{\frac{4}{3}}-2\omega^2  \right)^
{\left(\frac{\gamma+1}{4\left(\gamma-1\right)}\right)}
\right)^{-3}
\label{raf2}
\end{eqnarray}
\begin{equation}
{\frac {\rho_{f}}{\rho_{A}}} =
\left(1+\frac{\omega^{2/3}}{(\frac {r_{A}}{r_{f}})^2}
\left(1-(\frac {r_{A}}{r_{f}})^2\right)^{2/3}\right)^{-1} . 
\label{rhofa2}\end{equation}
If $\beta$ approaches zero, the fast point
is rejected to infinity, a well known result
in the cold plasma limit (Kennel et al. (\cite{kennel})) 
The dimensionless  specific energy $\epsilon$ can be 
derived by substituting these expressions
for the positions and densities at the critical points 
in the Bernoulli equation. We obtain: 
\begin{equation}
\epsilon = 3 \omega^{4\over 3} -
\frac{r_A}{r_f}\left(\omega^2
\left(1+\frac{4}{\omega^{{2}/{3}}}-
\frac{3}{\omega^{{4}/{3}}}\right)
+6\omega^{{2}/{3}}\right)
\label{Epsifastlong}
\end{equation}
and if $r_f \gg r_A$ it can be reduced to 
\begin{equation}
\epsilon = 3 \omega^{4\over 3} . 
\label{Epsifast}
\end{equation}
This gives the specific energy $E$ as a
function of the density and position
of the Alfv\'en point as:
\begin{equation}
E = {\frac {3^3 A^2}{2^4 \rho_A r_A^4}} .
\label{Efast}\end{equation}
Using the definitions of 
$\omega$ and $g$ which in the fast rotator 
limit can be written as
$g = 2 \omega^2 ({{r_{s}}/{r_{A}}})^3 $,
the position of the slow and the fast points 
are found to be given by
\begin{equation}
r_s = \left( \frac{G M}{\Omega^2 \cos{\theta}^2 }\right)^{1/3} .
\label{rsfast}\end{equation}
As the rotation increases, the slow point gets closer to the
source in proportion to $\Omega^{-2/3}$.
The fast point position is
\begin{equation}
r_f = \frac{\Omega^2 r_A^4 \cos{\theta}^2  }{G M} .
\label{rffast}\end{equation}
since $r_A$ will be found to increase with the rotation rate,
this expression shows that
since $r_A$ will be found to increase with the rotation rate,
the fast point is rejected far from the 
Alfv\'en point when the rotation grows very large.
Interesting physical consequences will be given in
the final discussion part.

%%%%%%%%%%%%%%%%%%%%%%%%%%%%%%%%%%%%%%%%%%%%%%%%%%%%%%%%%%%%%
\subsubsection{Fast cold rotator}
%%%%%%%%%%%%%%%%%%%%%%%%%%%%%%%%%%%%%%%%%%%%%%%%%%%%%%%%%%%%%

If moreover the wind is cold,
the fast point goes to infinity, in the limit of vanishing
entropy.
\begin{equation}  
\lim _{\beta \to 0} {\rho_f} =0
\label{limrhof}
\end{equation}
\begin{equation}  
\lim _{\beta \to 0} {r_f} =\infty
\label{limrf}
\end{equation}
It could be feared that the geometry 
of magnetic surfaces upstream from the fast point
 might not be conical at 
very high rotation rates. 
Numerical exploration of the
positions of critical points in models 
with  non-conical geometry 
having a cylindrical asymptotic shape has however shown
that, provided the rotation
does not grow to extreme values,
the fast point still remains in 
the quasi-conical region surrounding the source
and does not shift to the asymptotically 
cylindrical region.
Therefore our simple assumption may have 
a somewhat broader domain of validity 
than would be anticipated. This is why we felt
that it is still worth pursuing the study of
the approximate model even in this
limit where its validity becomes disputable.
When $\beta $ 
and $g$ are small compared to unity, 
we find
\begin{equation}
\omega \rightarrow 
\left(
\frac{3}{2}
\right)^{\frac{3}{2}} .
\label{omegaMAX}
\end{equation}
The Alfv\'en slope, for the fast rotator regime,
approaches a constant value
\begin{equation}
 p \rightarrow 1-\sqrt{\frac{27}{19}} .
\label{pLENT}\end{equation}
We can also deduce the radius and the velocity
at the Alfv\'en point as well as the specific energy
and angular momentum
\begin{equation}
R_A = 
\left(\frac{3}{2}\right)^{1/2}
\left(\frac{A}{\mu_0\alpha\Omega cos\theta}\right)^{1/3}
\label{Raveryfast}
\end{equation}
\begin{equation}
v_{PA} = 
\frac{A}{\mu_0 \alpha R_A^2} 
= \frac{2}{3}
\left(\frac{A \Omega^2 \cos^2\theta}
{\mu_0 \alpha} \right)^{1/3}
\label{vpaveryfast}
\end{equation}
\begin{equation}
E = \frac{27}{8} v_{PA}^2
= \frac{3}{2}
\left(\frac{A \Omega^2 \cos^2\theta}
{\mu_0 \alpha} \right)^{2/3}
\label{Everyfast}
\end{equation}
\begin{equation}
L = 
\frac{3}{2}
\left(\frac{A \Omega^{1/2} \cos^2\theta}
{\mu_0 \alpha} \right)^{2/3} .
\label{Lveryfast}
\end{equation}
we thus obtain expressions of 
the integrals of the motion
$E$ and $L$ relevant to this case. 
Inserting these results in the 
Alfv\'en regularity condition
 and integrating it, we find
\begin{equation} 
R_A^{\frac{16}{27}\sqrt{\frac{27}{19}}}
\left(\frac{cos\theta}{R_A}\right)
^{2(1-\sqrt{\frac{19}{27}})}
\frac {E\alpha}{\Omega}
= C_2
\label{Raveryfast2}
\end{equation}
where $C_2$ is an integration constant.
This gives the Alfv\'en radius in the form
\begin{equation}
R_A =
\biggl(\frac{C_2 \Omega}
{E \alpha (\cos\theta)^{2(1-\sqrt{\frac{19}{17}})}}
\biggr)
^{\frac
{\sqrt{513}}
{24-2\sqrt{513}}} .
\label{RadefFAST}\end{equation}
Comparing Eq.(\ref{Raveryfast}) 
and Eq.(\ref{RadefFAST}),
we find that the constant $C_2$
is related to other quantities by
\begin{equation}
C_2 = 
\alpha \Omega
\biggl(
\left(\frac{3}{2}\right)^{3/2}
\frac{A}{\mu_0\alpha\Omega cos\theta}
\biggr)^
{\frac{6}{\sqrt{57}}}
\cos\theta^{4(2-\sqrt{\frac{19}{17}})} .
\label{C2}
\end{equation}
When $\theta=0$, it becomes
\begin{equation}
C_2 = 
\alpha_0 \Omega_0
\biggl(
\left(\frac{3}{2}\right)^{3/2}
\frac{A}{\mu_0\alpha_0\Omega_0}
\biggr)^
{\frac{6}{\sqrt{57}}} 
\label{C2bis}
\end{equation}
and 
\begin{equation}
\frac {E\alpha}{\Omega} = 
\frac{3}{2} \left(
\frac{\alpha\Omega A^2}{\mu_0^2}\right)^{1/3} .
\end{equation}    
This last relation is directly related to the
escaping poloidal electric current.
So we have obtained a simple expression for the net current.
We have postulated that $\beta \ll 1$ in this part.
This is express by the 
following inequality :
\begin{equation}
\alpha \ll
\biggl( 
\left(\frac{2}{3}\right)^2
\frac{\gamma-1}{2 \gamma Q}
\frac{(A\Omega^2\cos\theta^2)^{2/3}}
{\mu_0^{{3\gamma-1}/3}}
\biggr)^
{\frac{3}{2(3\gamma-2)}}.
\end{equation}  
It shows that 
in the limit of our assumptions
$\alpha$ is restricted for
given value of all the other input 
parameters. We will come to this 
result later in the section dealing
with numerical results.

For very fast rotators, i.e. $\omega
 \sim \left(\frac{3}{2}
\right)^{\frac{3}{2}}$,
 we can easily express
this constant of integration, which then allows to
calculate all the variables and integrals 
of motion of the system. In this case 
the Eq.(\ref{C2}) closes the set of equations
that define the solutions, as did Eq.(\ref{C1def}) 
in the case of no rotation.

%%%%%%%%%%%%%%%%%%%%%%%%%%%%%%%%%%%%%%%%%%%%%%%%%%%%%%%%%%%%%
%\subsection{Constant Rotation and Entropy}
%%%%%%%%%%%%%%%%%%%%%%%%%%%%%%%%%%%%%%%%%%%%%%%%%%%%%%%%%%%%%

%%%%%%%%%%%%%%%%%%%%%%%%%%%%%%%%%%%%%%%%%%%%%%%%%%%%%%%%%%%%%
\subsubsection{An example of TTauri star: BP Tau}
%%%%%%%%%%%%%%%%%%%%%%%%%%%%%%%%%%%%%%%%%%%%%%%%%%%%%%%%%%%%%

This section is designed to show by an example the
behavior of the six variables for which our model
gives solutions in terms of boundary conditions 
and of the deduced quantities. 
As typical, we present the results for a typical TTauri star, BP Tau.

%%%%%%%%%%%%%%%%%%%%%%%%%%%%%%%%%%%%
\begin{figure}
%\picplace{6cm}
\psfig{figure=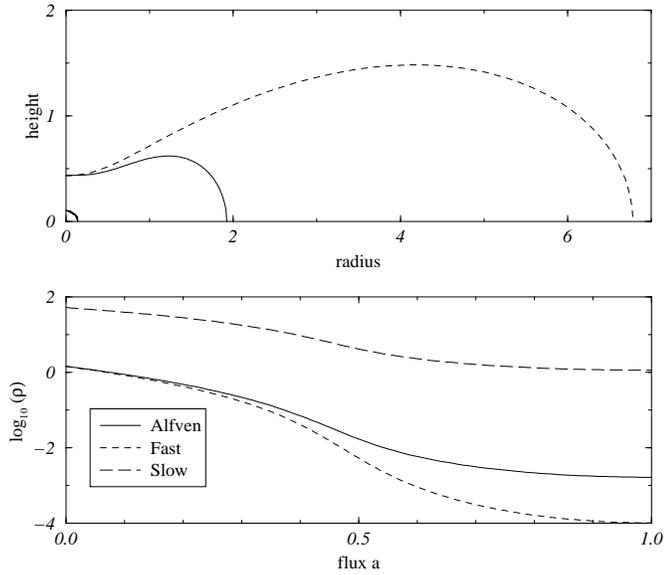,width=\linewidth}
\caption[ ]{
Example of the results obtained for a typical TTauri star
defined by $\bar{Q}=0.05$, $\bar{\Omega}=1.8$ 
and $\bar{\alpha}=0.1$ as input 
parameters ($\gamma = 1.2$). In the upper graph, the densities
 at the three critical points are represented with respect to the
magnetic flux a in linear-logarithmic frame.
The lower graph shows the corresponding critical surfaces
in dimension-less quantities. The Alfv\`en surface is the solid line
curve and the fast point surface the dashed line one, while
the slow surface has a heavy solid line and is situated
close to the origin. Note that, on the whole,
 the slow surface is almost circular
while others are not at all. 
}
\label{figEXAMPLE}
\end{figure}
%%%%%%%%%%%%%%%%%%%%%%%%%%%%%%%%

%%%%%%%%%%%%%%%%%%%%%%%%%%%%%%%%%%%%
\begin{figure}[htbp]
%\picplace{5.5cm}
\psfig{figure=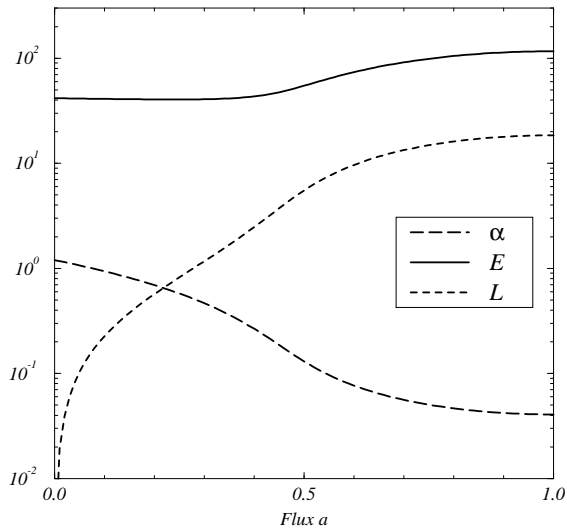,width=\linewidth}
\caption[ ]{Example of variations of the three first
integrals of the motion for a typical TTauri star
(BP TAU, Bertout et al. (\cite{bertout}))
 with respect to the magnetic flux a.
they are deduced from the positions
and densities of the critical points.
The solid line represents the specific energy $E$,
the dashed line the angular momentum $L$ and
the long-dashed line the mass loss to magnetic flux
ratio $\alpha$.
}
\label{figFIRSTINT}
\end{figure}
%%%%%%%%%%%%%%%%%%%%%%%%%%%%

From Bertout et al. (\cite{bertout}), we take the reference values as
$\dot M_{\ast}=2\times10^{-7} M_{\odot} yr^{-1}$,
$M_{\ast}=0.8 M_{\odot} $,
$R_{\ast}=3 R_{\odot}$,
$T_{\ast}=9\times10^3$,
$n_p=10^4 cm^{-3}$,
and $B{\ast}=1000 G$.
We deduce the dimensionless input parameters 
${\overline Q} =0.05$, 
${\overline \Omega} =1.8$ 
and ${\overline \alpha_0} =0.1$.
 The three critical surfaces (upper panel) and corresponding 
densities (lower panel) are represented
 on Fig. \ref{figEXAMPLE}.
The lower part of the figure shows clearly the 
decreasing trend of the densities from the axis  
to the equator and stresses the differences in 
magnitude of the densities at the different critical points, 
the density at the slow point being one hundred 
times larger than the others. This reflects the 
position of the critical surfaces. 
On the upper part
of figure 2, it can be seen
that the fast and Alfv\'en critical surfaces 
definitely exhibit  shapes different from spherical,
particularly the fast one. One can notice that
the further we go away from the axis
the more the fast and the Alfv\'en surfaces get apart.
%************************************************************************
%REFEREE*************************************************************
As seen in Sakurai (\cite{saku2}) and Belcher 
\& McGregor (\cite{belcher}), as one gets closer 
to the equator the critical surfaces get more and more  
elongated along the equator. However a difference appears
near the axis of rotation. In the latter cases, 
the critical surfaces get elongated along the pole.
This is mainly due to the constancy of the angular velocity
$\Omega(a)$ and of the entropy $Q(a)$. When this assumption
is relaxed, equivalent behaviors have be obtained.  
On the other hand rotation has little effect on the 
slow mode critical surface which almost  keeps 
a spherical shape. 
In fact the slow surface gets closer as the velocity increases,
as one would expect for such a type of wave considering the 
increase of centrifugal acceleration near the equator. 
The behavior of this surface will be studied  in
more details in next sections.
%REFEREE **************************************************************
%************************************************************************
From the positions and densities 
of the critical points we deduce 
the other interesting variables of the problem, such as
the first integrals of the motion $E$, $L$ and $\alpha$ 
as represented on Fig. \ref{figFIRSTINT}.
The specific energy  seems to vary little as
compared to the other quantities.
Its variation is less than one percent of the mean value.
Thereafter the amount of energy is  slowly increasing from
the value on the axis, given by Eq.(\ref{Es}),
to an higher value on the equator.
The specific angular momentum
increases from the axis and then reduces its 
growth to level off at a value close to unity.
The behavior of $\alpha$ is identical to that of
the Alfv\'enic density as seen in Eq.(\ref{rhoa}).
One should keep in mind that the value of 
$\alpha$ on the axis is prescribed as an input 
parameter.

In this subsection, we have shown 
a particular set of numerical solutions 
of the equations of this model.
In the following sections we comment on the 
influences of the input parameters on the solutions.

%%%%%%%%%%%%%%%%%%%%%%%%%%%%%%%%%%%%%%%%%%%%%%%%%%%%%%%%%%%%%
\subsubsection{Effects of the amplitude of the Mass Loss}
%%%%%%%%%%%%%%%%%%%%%%%%%%%%%%%%%%%%%%%%%%%%%%%%%%%%%%%%%%%%%

%%%%%%%%%%%%%%%%%%%%%%%%%%%%%%%%%%%%
\begin{figure}[htbp]
%\picplace{3.7cm}
\psfig{figure=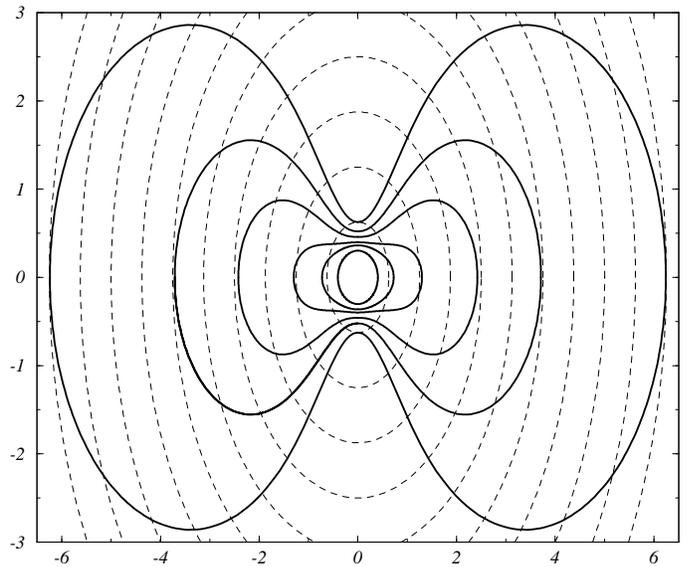,width=\linewidth}
\caption[ ]{
The Alfv\'en surfaces for increasing $\alpha_0$
are shown with solid lines.
Circular shapes are represented with dashed lines
for comparisons with the solutions. Smaller surfaces 
correspond to higher 
values of  $\alpha_0$ while most distorted surfaces 
have smaller $\alpha_0$.
 The smaller $\alpha_0$, the bigger the distortion
and the higher the height of the Alfv\`en surface
on the polar axis.
Small $\alpha_0$ can either correspond to low mass loss rate, or to 
high magnetic flux.
}
\label{figAlfALPHA}
\end{figure}
%%%%%%%%%%%%%%%%%%%%%%%%%%%%%%%%%%%%
%%%%%%%%%%%%%%%%%%%%%%%%%%%%%%%%%%%%
\begin{figure}[htbp]
%\picplace{5cm}
\psfig{figure=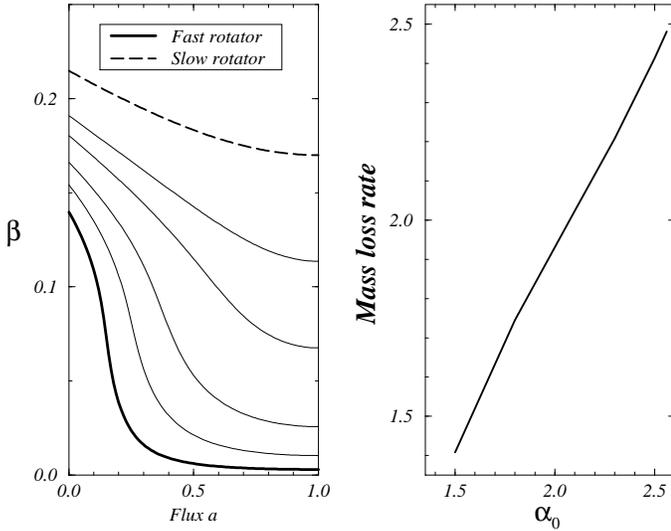,width=\linewidth}
\caption[ ]{
As shown on the left panel, $\beta$ is small compared
to one for various values of $\alpha_0$. The upper long-dashed
curve corresponds to high $\alpha_0$ while going down the curves,
one finds solutions for smaller ratios. Those curves strengthen
the assumption for our analytical calculations where $\beta$
is considered small compared to one.
On the right panel, the curve indicates
the total mass loss rate against $\alpha_0$.
Notice that the slope of the curve is almost unity.
}
\label{figMdotalpha0}
\end{figure}
%%%%%%%%%%%%%%%%%%%%%%%%%%%%%%%%%%%%

The mass loss rate $\dot{M}$ is a quantity that 
can be deduced from observations. 
In our model, it is represented by
the parameter $\alpha_0$, the mass flux to 
magnetic flux ratio on the polar axis. The relation
between $\alpha_0$ and $\dot{M}$ 
can be worked out a posteriori. We have 
calculated solutions for different $\alpha_0$'s,
keeping for the other parameters the same 
value as in the previous calculations.
The corresponding Alfv\`enic surfaces are presented in 
Fig. \ref{figAlfALPHA}. Smaller
surfaces correspond to high $\alpha_0$'s while the 
distorted ones correspond to smaller values. 
This variation is similar for all the critical surfaces. 
The critical surfaces strongly inflate for decreasing 
$\alpha_0$ and lose the initially spherical shape that 
they have at high mass loss rates.
It shows that for strongly magnetized systems,
the structure of the zone of acceleration gets complex.

Fig. \ref{figMdotalpha0} shows 
the evolution of $\beta$ as a function of the magnetic flux
for different $\alpha_0$'s. Fig. \ref{figMdotalpha0}
shows the simple relation which exists  between
 $\alpha_0$ and the  mass loss rate with the other input
 parameters fixed. In Fig. \ref{figMdotalpha0} the values 
of $\beta$, given by Eq.(\ref{beta}),
are represented against the relative magnetic flux
for a series of  decreasing values of $\alpha_0$. 
When $\alpha_0$ is large, the behavior of $\beta$ is
nearly linear but for lower values 
the curves start to
decrease steeply and reach a low equatorial value.
This strengthens one of the assumptions made
to derive our analytical results, namely that
the value of $\beta$ is small compared to unity.
The value of $\beta$ on the axis depends directly
on $Q$. If the entropy is large enough, $\beta$ can become
large too and even reach unity. This would be 
the case for very hot outflows. The smaller $\alpha_0$ ,
the better the assumption of smallness of $\beta$ is justified. 
Note that the critical surfaces for magnetized
 rotators with small mass loss rates are strongly 
inflated equator-wards
and are in no way similar to the quasi-spherical shape 
they have for rotators with large $\dot M$.
In this section we have shown that
fast and slow rotators developed different shapes
of the critical surfaces.
It does not seem possible to 
infer the solution for the fast rotator 
from the one obtained for a slow rotator by 
simple scaling scaling arguments
as attempted by Shu et al. (\cite{shu94}).

%%%%%%%%%%%%%%%%%%%%%%%%%%%%%%%%%%%%%%%%%%%%%%%%%%%%%%%%%%%%%
\subsubsection{Rotational effects}
%%%%%%%%%%%%%%%%%%%%%%%%%%%%%%%%%%%%%%%%%%%%%%%%%%%%%%%%%%%%%

%%%%%%%%%%%%%%%%%%%%%%%%%%%%%%%%%%%%
\begin{figure}[htbp]
%\picplace{2.3cm}
\psfig{figure=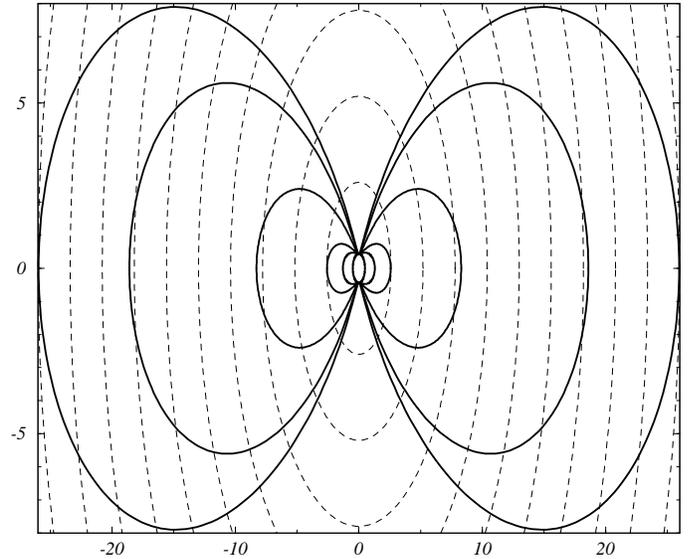,width=\linewidth}
\caption[ ]{
Effects  of the angular velocity $\Omega$ on
fast surfaces with parameters of  a typical TTauri star.
 Circles are shown to visualize
the distortions of the critical surfaces.
These curves indicates that the faster the
rotation, the bigger the distortion with
respect to sphericity is.
The position of the critical points on the polar axis
is of course independent of the rotation.
}
\label{figFASTsurfOM}
\end{figure}
%%%%%%%%%%%%%%%%%%%%%%%%%%%%%%%%%%%%
\begin{figure}[htbp]
%\picplace{5cm}
\psfig{figure=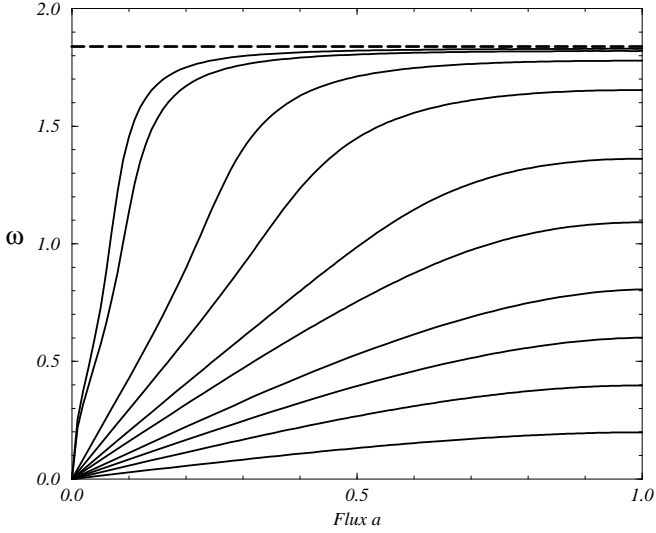,width=\linewidth}
\caption[ ]{
The variation of the rotation parameter $\omega$ 
with to the magnetic flux $a$
is plotted
for different values of the rotation rate $\Omega$ .
The asymptotic limit of  $(\frac{3}{2})^{\frac{3}{2}}$ 
is represented with an heavy dashed line. 
Lower curves represent slow rotators while upper curves stand
for fast rotators. An arbitrary limit separate the two type of
rotators at $\omega \sim 1$. 
Close to the axis, a slow rotator type behavior always remain
whatever the rotation rate.
}
\label{figOMom}
\end{figure}
%%%%%%%%%%%%%%%%%%%%%%%%%%%%%%%%%%%
\begin{figure}[htbp]
%\picplace{5cm}
\psfig{figure=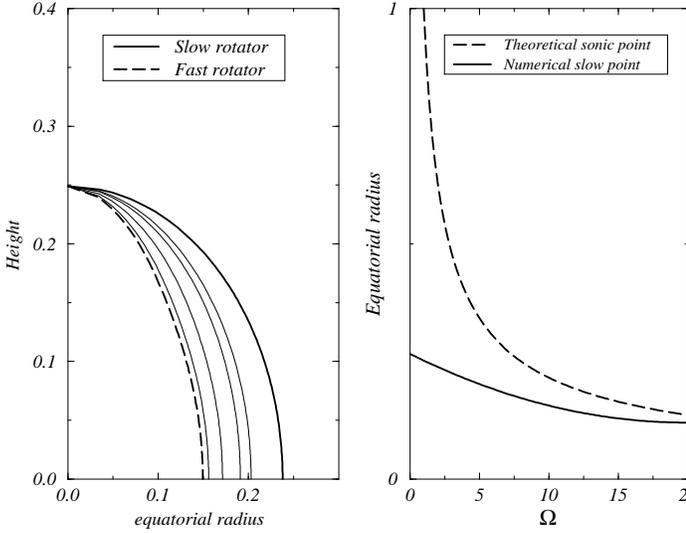,width=\linewidth}
\caption[]{
The shapes of the slow magneto-sonic surfaces for increasing rotation rates
$\Omega$ are shown on the left panel. The slow surface has a opposite
behavior compared to the fast and the Alfv\`en surfaces. It deflates while
$\Omega$ increases. 
On the right panel  of the figure, the numerical solution for the 
magnetosonic slow point is  represented with the 
variations of the pure sonic point with respect
to the rotation rate. This shows that the slow magnetosonic point
tends to the trend of the sonic point for increasing rotations.
}
\label{figrsequateur}
\end{figure}
%%%%%%%%%%%%%%%%%%%%%%%%%%%%%%%%%%%%

The value of the angular velocity has a major influence
on the behaviors of the solution. In Fig. \ref{figFASTsurfOM}
several  Alfv\'enic surfaces are represented for
different $\Omega$.
One should notice that the central emitting
object is reduced to a point and does not coincide with  the 
first circular surface.
This innermost surface corresponds to the
smallest rotation rate and the other surfaces 
expand off the central part for larger and larger 
rotation rates.
The faster 
the rotator, the larger the change in  the 
shape of  alfv\'enic surfaces.
The effect of the rotation is strongest at 
the equator. The Alfv\'en surface remains at a 
finite distance from the source for finite rotation rates. 
The Alfv\'en point only goes 
asymptotically to infinity for ultra-fast rotators.
 
In Fig. \ref{figOMom}
we present a plot of  $\omega(a)$ for different 
rotation speeds
 with the same previous input parameters 
$Q$ and $\alpha_0$.
 The lower curves illustrates slow rotation rates
and those above them correspond to increasingly 
larger rotation.
The horizontal dashed lines represents the asymptotic
limit for ultra-fast rotators where $\omega={3/2}^{3/2} 
\approx 1.837$. The curves pass smoothly from
the region of slow to fast rotation.
Once the asymptotic value for $\omega$ is nearly reached
on the equator, the asymptotic region extends 
towards the axis for larger rotation reducing the 
extent of the region  where the rotation parameter 
stays small. This region region never disappears
completely, though, because obviously the polar axis 
itself is necessarily in the slow rotation regime.

For fast rotators
the Alfv\`en point is at a large distance. In this case, 
we see that the alfv\'en speed is large compared to the 
sound speed. At the equator the slow magneto-sonic 
surface as seen in Fig. \ref{figrsequateur}  gets closer 
to the source as the rotation rate increases.
%************************************************************************
%REFEREE% REFEREE ***************************************
The effect of the centrifugal acceleration
is more and more pronounced as one reaches
the equator.
%REFEREE% REFEREE ***************************************
%************************************************************************
The slow mode speed  gets  closer to the sound 
speed, and the slow mode acquires  the character of 
a sound wave  guided along the field line. For this 
reason we can see  the slow point on the equator 
moving to the source. The radius of the slow mode 
surface on the equator approaches the radius where
the sound speed corotates with the Keplerian speed,
given by:
\begin{equation}
r_s = \biggl(\frac{GM}{\Omega^2}\biggr)^{1/3}  
\end{equation}
We have represented the radius at the slow point
situated on the equatorial plane as a function of
increasing rotation speeds 
on the right panel of Fig. \ref{figrsequateur}
together with the pure sonic point.
It is clear that the slow magneto-sonic point 
tends to the sonic point.
We now have a view of how 
all the critical surfaces 
behave as the rotation rate vary.
Fast magneto-sonic and Alfv\`en surfaces strongly inflate 
when rotation increases while the slow surface gets
narrower to converge to a nearly  cylindrical shape.
It seems that an increase of $\Omega$ has almost the
same effects as a decrease of $\alpha$, except that
the positions of the critical points do not change 
on the polar axis when the rotation rate varies.

%%%%%%%%%%%%%%%%%%%%%%%%%%%%%%%%%%%%%%%%%%%%%%%%%%%%%%%%%%%%%
\subsubsection{Thermal effects}
%%%%%%%%%%%%%%%%%%%%%%%%%%%%%%%%%%%%%%%%%%%%%%%%%%%%%%%%%%%%%

%%%%%%%%%%%%%%%%%%%%%%%%%%%%%%%%%%%%
\begin{figure}[htbp]
%%\picplace{8cm}
\psfig{figure=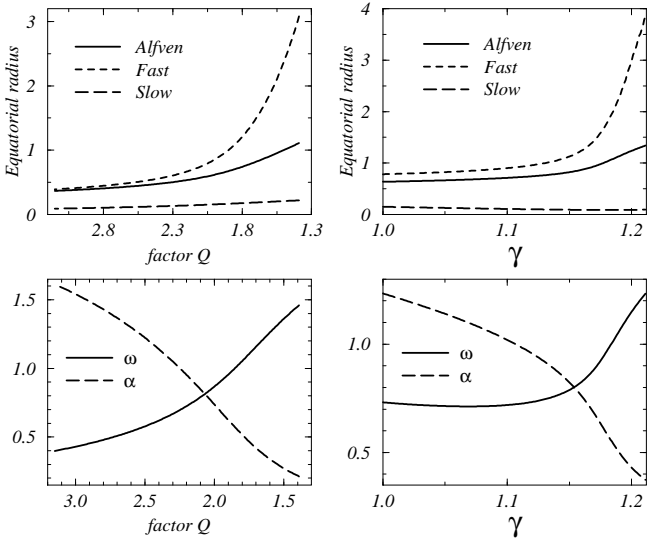,width=\linewidth}
\caption[ ]{
Some of the thermal effects are represented on this figure.
The upper panels show the variations with Q and $\gamma$ 
of the distance to the origin of
the critical points in the equatorial plan.
Below we show how
 $\omega$ and $\alpha$ vary. smaller entropies  make
 the rotation parameter smaller 
while they produce larger ratios of mass loss
rate to magnetic flux (Note that the entropy
decreases from left to right). 
An increase of $\gamma$ have the same
properties. The isothermal
case produces the larger mass loss rate
and as $\gamma$ increases the rotator
turns to be faster.
}
\label{figGAMMAQ}
\end{figure}
%%%%%%%%%%%%%%%%%%%%%%%%%%%%%%%%%%
We have assumed a polytropic equation of state.
Thus the two parameters that we can vary 
to illustrate the thermal effects are the entropy 
and the polytropic index. We consider these 
parameters as  constants as a function of flux 
variable $a$. In these calculations, $\Omega$ and 
$\alpha_0$ are fixed.
 When Q, the factor related to the entropy,
decreases, the critical surfaces inflate, 
the rotation parameter $\omega$
 increases and $\alpha$ reduces. This can be understood
as follows: 
the more the plasma is heated, 
the lesser is the influence of the rotation.
When the polytropic index is varied and the thermodynamics
turns progressively from isothermal to adiabatic,
the influence of rotation becomes more and more important.
This is because the effect of gas pressure 
on the dynamics is 
maximized by an infinite thermal conduction. 
For most solutions
the ratio of the fast mode radius
to the Alfv\`en radius is not very large.
Thus the variations of $Q$ and $\gamma$ have a 
non-negligible influence but nevertheless weaker 
than that of varying 
$\Omega$ and $\alpha_0$.
A reduction of the entropy has a similar effect to that 
of an increase of the polytropic index $\gamma$.
 The smaller the entropy, the closer 
the flow becomes to that of a fast rotator.

%%%%%%%%%%%%%%%%%%%%%%%%%%%%%%%%%%%%%%%%%%%%%%%%%%%%%%%%%%%%%
\subsubsection{Limit cases}
%%%%%%%%%%%%%%%%%%%%%%%%%%%%%%%%%%%%%%%%%%%%%%%%%%%%%%%%%%%%%

In the previous section, we have seen that
the fast rotator regime is achieved before $Q$ 
vanishes. Similarly, all the other parameters 
vary in a restricted range.
For example,as the rotation $\Omega$ increases,
the rotation parameter $\omega$ 
reaches the maximum value 
$(\frac{3}{2})^{\frac{3}{2}}$.
Numerically we find that there is  an lower limit to 
$\alpha_0$ for given  $\Omega$ and other parameters. 
To understand it, let us recall the equation
$ g = 2 G M \mu_0^2 \alpha^2 R_A^3 A^{-2}$.

It has been found that
\begin{equation}
\omega = \frac{\Omega cos \theta \mu_0 \alpha R_A^3}{A} .
\end{equation}
So for fast rotators the Alfv\`enic radius tends to be
almost equal to
\begin{equation}
R_A =  \left(\frac{3}{2}\right)^{1/2}
\left(\frac{A}{\mu_0 cos \theta}\right)^{1/3}
(\alpha \Omega)^{1/3}
\end{equation}
The gravity parameter the becomes
\begin{equation}
g = \frac{2 G M}{r_A vp_A} =
\left(\frac{3}{2}\right)^{3/2}
\left(\frac{2 G M \mu_0}{A cos \theta^2}\right)
\frac{\alpha}{\Omega}
\end{equation}
Meanwhile we have found that
this parameter could be expressed as
\begin{equation}
g = 2 \omega^2 x_s^3 = \frac{27}{4} x_s^3.
\end{equation}
It shows that $g$ is small with respect to unity
and that it does not vary much since $x_s$
mainly depends on the thermal parameters and not on
much on the mass loss rate or the rotation rate
in the case of fast rotators.
It also shows that in the case of very fast rotators
either the angular velocity must be very important or
the mass loss rate must be small to insure that $g$
is small.
One can  also deduce that there is a maximum 
$\Omega$ for a given such as
$\alpha$ 
\begin{equation}
\Omega_{max} =  \omega_{max} f(\alpha_0,Q)
\label{Omegamax}
\end{equation}
The fitting of the numerical limits of our model
where 
$\omega = \left(\frac{3}{2}\right)^{\frac{3}{2}}$ 
agrees with this result and gives
\begin{equation}
\Omega_{max} = \left(\frac{3}{2}\right)
^{\frac{3}{2}}  \left(1+\frac{Q}{2}\right) 
 \alpha_0^{\frac{2}{3}}
\label{Omegamax2}
\end{equation}
Thus this shows that if the outflow is 
to support a given mass loss rate, it cannot
rotate at any large rate. 
The converse is also true:
if the magnetic field lines
turn at a given rotation rate $\Omega$, 
there is a minimum  mass loss rate from the 
central object.

%%%%%%%%%%%%%%%%%%%%%%%%%%%%%%%%%%%%%%%%%%%%%%%%%%%%%%%%%%%%%
\subsubsection{Discussion}
%%%%%%%%%%%%%%%%%%%%%%%%%%%%%%%%%%%%%%%%%%%%%%%%%%%%%%%%%%%%%

We have presented an extended set 
of solutions of the model, 
for constant $\Omega$ and $Q$ across the magnetic field lines.
We obtain the first integrals of the motion,
the components of the velocity, of the  magnetic field
and all the variables characterizing the flow.
This study revealed some interesting properties. 
We have shown that an increase of $\Omega$
up to the fast rotator regime has a similar effect to 
a decrease of $\alpha_0$, or $Q$, and to an
increase of $\gamma$. Fast rotators display  
critical surfaces which are strongly inflated 
equator-wards, that drastically differ from
their quasi-spherical shape in the slow rotator regime.
The faster the rotator, the more valid the assumption 
that $\beta$ is small. The influence of thermal 
parameters is smaller than that of rotation or mass 
loss rate but not negligible. Finally we have also found
that a given rotation rate rate compels the mass loss rate 
of the outflow to have an inferior limit.

%%%%%%%%%%%%%%%%%%%%%%%%%%%%%%%%%%%%%%%%%%%%%%%%%%%%%%%%%%%%%
\subsection{Other variations of $\Omega$ and $Q$}
%%%%%%%%%%%%%%%%%%%%%%%%%%%%%%%%%%%%%%%%%%%%%%%%%%%%%%%%%%%%%

%%%%%%%%%%%%%%%%%%%%%%%%%%%%%%%%%%%%
\begin{figure}[htbp]
%\picplace{6cm}
\psfig{figure=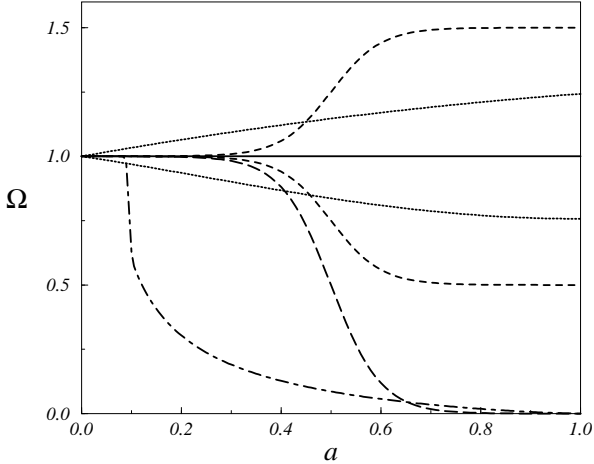,width=\linewidth}
\caption[ ]{
The variations of the rotation rate $\Omega$ with $a$
for the differential rotators. considered in this work.
The dot-dashed line represents a Keplerian rotation.
After from the bottom, one find a step-shape rotation
with $\Omega$ varying from 1 to 0, a step-shape 
rotation from 1 to 0.5,
a decreasing stellar type rotation,a constant one,
a normal stellar type differential rotation 
and finally a step-shape
rotation from 1 to 1.5. 
}
\label{figFORMomega}
\end{figure}
%%%%%%%%%%%%%%%%%%%%%%%%%%%%%%%%%%%%

The distribution of outflowing material in space 
and velocity is not well determined observationally, 
but some properties of such flows have still be recognized.
The relatively wide spectral lines which have been 
measured at most places in such outflows 
indicate that material with a range of velocities is 
present on the line of sight. However, in a number 
of well-collimated outflows, such as NGC2024 and 
NGC2264G, there appears to be a shear flow, with 
higher velocities near the polar axis, and lower 
velocities at the sides of the flow (Margulis et al. (\cite{margu}); 
Richer et al. (\cite{RHP})). These observations motivated us 
to  look for outflows with rotation rates and 
entropy varying with the flux variable $a$,
that could reproduce such behaviors.

%%%%%%%%%%%%%%%%%%%%%%%%%%%%%%%%%%%%
\begin{figure}[htbp]
%\picplace{7cm}
\psfig{figure=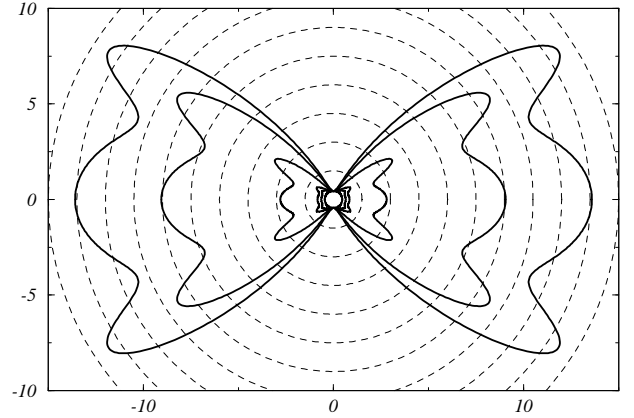,width=\linewidth}
\caption[ ]{
Fast critical surfaces for a differential rotator.
In this example the rotation
rate starts with a fast rotation on the axis 
and then reduces to a smaller rotation rate on the equator.
The surfaces start just like fast rotators shown on Fig.(\ref{figFASTsurfOM})
but change drastically their shapes afterwards giving this butterfly shape. 
}
\label{figVARomfastDIFF}
\end{figure}
%%%%%%%%%%%%%%%%%%%%%%%%%%%%%%%%%%%%

%%%%%%%%%%%%%%%%%%%%%%%%%%%%%%%%%%%%%%%%%%%%%%%%%%%%%%%%%%%%%
\subsubsection{Non-constant $\Omega$}
%%%%%%%%%%%%%%%%%%%%%%%%%%%%%%%%%%%%%%%%%%%%%%%%%%%%%%%%%%%%%

%%%%%%%%%%%%%%%%%%%%%%%%%%%%%%%%%%%%
%\begin{figure}[htbp]
%\picplace{7cm}
%\psfig{figure=graph3D.ps,height=7truecm,width=8truecm}
%\caption[ ]{
%Three-dimensional representation of
%a fast critical surface for a differential rotator.
%The full shape of the surface can be visualized.
%This stresses the fact that spherical shapes
% are far too bad approximations for critical surfaces.
%}
%\label{figgraph3D}
%\end{figure}
%%%%%%%%%%%%%%%%%%%%%%%%%%%%%%%%%%%%

We now describe some specific forms of the rotation 
rates and discuss trends and properties that can be identified.
For the rigid body rotation, the angular velocity is constant
with flux $a$. The stellar-type rotation is inspired
by the equation of differential rotation of the sun.
We have modeled a stellar-type rotation with angular velocities
larger at the equator than at the pole.
We have also formulated variations of the rotation rate
that could account for two components outflows such as the models of
Shu et al (\cite{shu94}).
A central fast flow is surrounded by a slower wind.
We present two different differential rotators with
this characteristics. We call them two-velocities outflows with 
a step-shape rotation. The analytical expressions of all these 
differential rotation rates can be found in appendix C.
We have plotted them as functions of the magnetic flux
on Fig. \ref{figFORMomega}.
Solutions of the problem with such profiles of rotation 
differ from previous ones. Critical surfaces, densities, and 
all the other variables, strongly vary with $a$.
Fig. \ref{figVARomfastDIFF} is an illustration of
of the fast critical surfaces for a two-velocities outflow.
This produces ``butterfly'' shapes showing the possibility
for two kinds of outflow with different characteristic
to live side by side. 

\begin{figure}[htbp]
\psfig{figure=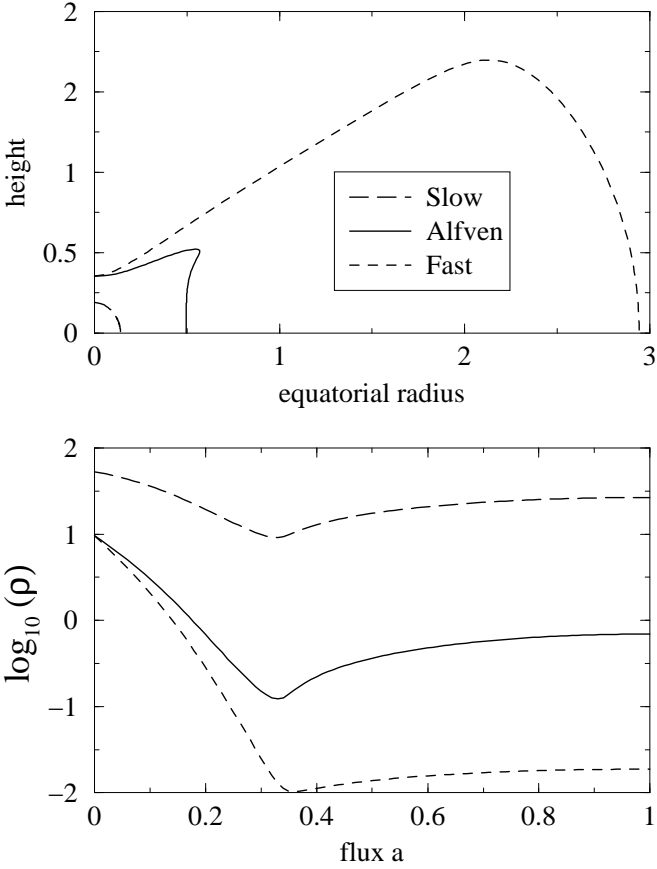,width=\linewidth}
\caption[ ]{
Positions and densities of the the slow, Alfv\'enic and
fast magnetosonic critical points for a fast 
rigid rotator surrounded by a keplerian disk.

}
\label{figKEPL}
\end{figure}

We have also plotted 
the solution for a central rigid rotator
surrounded by a keplerian disk in
Fig. \ref{figKEPL}. The central part
near the axis of rotation has got a constant
rotation velocity up to $a=0.3$, then
$\Omega$ follows a keplerian variation. 
This kind of model could apply to
Young Stellars Objects where the central
object is launching an O-wind
and the accretion disk an X-wind
(see Shu et al. (\cite{shu94})).

%%%%%%%%%%%%%%%%%%%%%%%%%%%%%%%%%%%%
%%%%%%%%%%%%%%%%%%%%%%%%%%%%%%%%%%%%

%%%%%%%%%%%%%%%%%%%%%%%%%%%%%%%%%%%%%%%%%%%%%%%%%%%%%%%%%%%%%
\subsubsection{Non-constant $Q$}
%%%%%%%%%%%%%%%%%%%%%%%%%%%%%%%%%%%%%%%%%%%%%%%%%%%%%%%%%%%%%

The variation of the entropy with magnetic flux $a$ seems to
have little influence on the solution 
in our computations, presumably because $\beta$ remained
small either due to small $Q$ values or to the rapid
rotator regime we imposed in several cases.
Nevertheless we have studied different possibilities and only in
in limit cases of the value of $Q$ and $\gamma$ the results change. 
In previous sections we have presented solutions for
constant entropy with different values. 
We have also tested different variations with $a$ of the function $Q(a)$,
with increasing or decreasing entropies from pole to equator and
a step-shape profile. Details can be found in appendix C.
The general trend is in line with the results
obtained in our study of thermal effects. For either increasing or 
decreasing entropies, critical surfaces respectively get closer
to the source or inflate outwards. For step shape variations
the critical surfaces are similar to surfaces represented in
Fig. \ref{figVARomfastDIFF}.

%%%%%%%%%%%%%%%%%%%%%%%%%%%%%%%%%%%%%%%%%%%%%%%%%%%%%%%%%%%%%
\section{Intermediate rotator}
%%%%%%%%%%%%%%%%%%%%%%%%%%%%%%%%%%%%%%%%%%%%%%%%%%%%%%%%%%%%%

%%%%%%%%%%%%%%%%%%%%%%%%%%%%%%%%%%%%%%%%%%%%%%%%%%%%%%%%%%%%%
\subsection{Analytical results}
%%%%%%%%%%%%%%%%%%%%%%%%%%%%%%%%%%%%%%%%%%%%%%%%%%%%%%%%%%%%%

Considering $\omega$ close to unity and 
neglecting $\beta$ with respect to unity
it is again possible to obtain analytical results
for the energy and gravitation parameters.
At the slow point we get
\begin{equation}
\epsilon = \beta y_s^{\gamma-1} -\frac{3}{x_s^4 y_s^2}
+2\omega^2
\end{equation}
and
\begin{equation}
g = 2 \omega^2 x_s^3.
\end{equation}
At the fast point the solution for these two parameters
is given by
\begin{equation}
\epsilon = 3\omega^{4/3}
\end{equation}
and 
\begin{equation}
g = \frac{2}{x_f}
\left(\omega^2+3\omega^{2/3}
-\frac{4}{3}\omega^{4/3}\right) .
\end{equation}
It is possible to deduce equations for the
positions and densities at the slow point
if one considers that 
$y_s \gg y_A$ and $x_s \ll x_A$ and gets
\begin{equation}
y_s = \left(
\frac{\beta(\gamma-1) x_s^4}{2(1+\omega^2 x_s^2)} 
\right)^{\frac{1}{1+\gamma}}
\end{equation}
\begin{equation}
\frac{r_s}{r_A} = 
\left(\frac{\beta(\gamma-1)}{2}\right)^
{\frac{1}{2(\gamma-1)}}
\left(\frac{5-3\gamma}
{(\gamma-1)(3\omega^{4/3}-2\omega^2)}\right)^
\frac{\gamma+1}{4(\gamma-1)}
\end{equation}
Using the last two equations one gets a better
equation for the density
\begin{equation}
\frac{\rho_s}{\rho_A} = 
\left(\frac{\beta(5-3\gamma)}
{2(3\omega^{4/3}-2\omega^2)}\right)^
{\frac{1}{\gamma-1}}
\end{equation}
It is also possible to get the variations of the
position and density at the fast critical point.
In the intermediate rotator regime, no direct
simplification can be done on the positions
of the critical surface, so the full set of solutions
is hard to find analytically but nevertheless
the solution is fully determined.
Only numerical studies can help to understand the behavior
 of such solutions in this regime. 
This is what the next part intends to do
with comparisons of the numerical 
and analytical results.

%%%%%%%%%%%%%%%%%%%%%%%%%%%%%%%%%%%%%%%%%%%%%%%%%%%%%%%%%%%%%
\subsection{Analytical vs numerical results}
%%%%%%%%%%%%%%%%%%%%%%%%%%%%%%%%%%%%%%%%%%%%%%%%%%%%%%%%%%%%%

%%%%%%%%%%%%%%%%%%%%%%%%%%%%
\begin{figure}[htbp]
%\picplace{6cm}
\psfig{figure=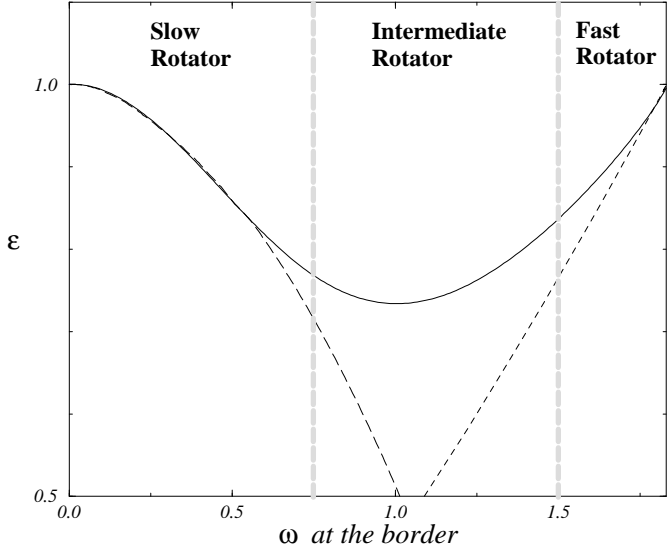,width=\linewidth}
\caption[ ]{
Comparisons of numerical results and analytical calculations
of the specific energy. On the left, $\epsilon$ is represented as 
a function of the magnetic rotator energy $\omega$.
There are good agreements for slow and  very fast rotators.
This is not the case of the intermediate region around 
$\omega=1$ that has no analytical solution.
But this region is really narrow with respect to the magnetic flux a 
as shown on the plot on the right hand side. 
}
\label{figcomp2}
\end{figure}
%%%%%%%%%%%%%%%%%%%%%%%%%%%

The analytical solutions and the numerical results 
 obtained for the specific energy and presented in the previous parts
have been plotted in Fig.(\ref{figcomp2}) for comparison. 
The figure shows the specific energy against the 
rotation parameter $\omega$. 
The results agree
in the slow regime and in the very fast regime,
as expected.
The intermediate zone that we have arbitrarily situated
between $\omega=0.75$ and $\omega=1.5$ shows clearly
the limit of validity of the solutions for
the slow and fast rotators.
It shows a smooth transition between the different
categories of rotators.
In fact, this region is narrow in the physical space
for a fast rotator since it is the transition between
the central slow part of the jet close the axis of rotation
and its fast part. So the analytical results are a good
approximation of the numerical solution.
Similar comparisons have been made for the positions
and densities of the critical points 
with similarly positive conclusions.

%%%%%%%%%%%%%%%%%%%%%%%%%%%%%%%%%%%%%%%%%%%%%%%%%%%%%%%%%%%%%
\section{Discussions}
%%%%%%%%%%%%%%%%%%%%%%%%%%%%%%%%%%%%%%%%%%%%%%%%%%%%%%%%%%%%%

%%%%%%%%%%%%%%%%%%%%%%%%%%%%%%%%%%%%%%%%%%%%%%%%%%%%%%%%%%%%%
\subsection{Comparison with other models}
%%%%%%%%%%%%%%%%%%%%%%%%%%%%%%%%%%%%%%%%%%%%%%%%%%%%%%%%%%%%%

\begin{figure}[htbp]
%\picplace{6cm}
\psfig{figure=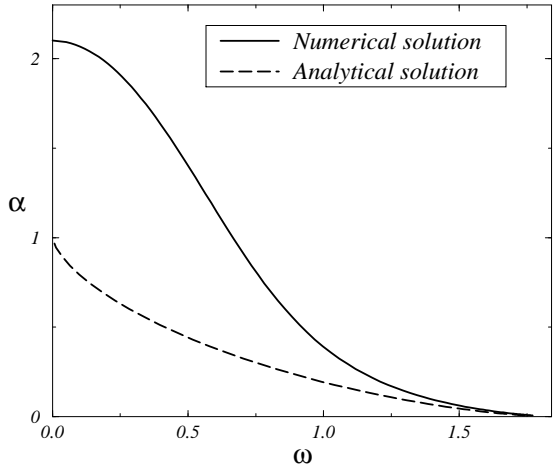,width=\linewidth}
\caption[ ]{
Comparisons of numerical results and analytical calculations
of the relation between $\omega$ and $\alpha$ on 
the equator. The upper curve (solid line) represents 
the numerical result while the lower one  (dashed line) 
the analytical solution as derive following the formalism 
of Spruit (\cite{spruit}). The entropy is taken as constant.
 The two curves converge to each other when
the rotation parameters becomes larger, which
 happens when $\alpha$ decreases. We find again 
that the bigger the mass loss rate the slower the 
rotation parameter.
}
\label{figSPRUIT}
\end{figure}

In a review of the theory of magnetically accelerated
 outflows and jets from accretion disks, Spruit (\cite{spruit})
discusses how the wind structure should depend on
mass flux (see also Cao \& Spruit (\cite{cao}) ). 
Using the model of  Weber \& Davis (\cite{wd}) he
concludes that when the mass flux is small, the Alfv\`en radius
extends far away from the origin. 
Our analytical and numerical results 
agree with these conclusions, with even more generality,
since our model fills all space and does not assume
any a-priori variations of the different variables.
We find, using the conditions of vanishing of the 
differential form of the Bernoulli equation,
that, if $\beta $ is small compared to unity, 
the Alfv\`en radius on the equator can be 
approximated by 
\begin{equation}
\frac{ R_A}{R_*} = \frac{5}{9} 
\left(\frac{\alpha}{\alpha_*}\right)^{-1/3}  .
\label{RAspruit}
\end{equation}
For large mass loss rate, Spruit finds that the Alfv\`en point
does not recede arbitrarily close to the origin  but reaches a
 minimum value. In our model when $\alpha$ becomes large
the fast point can be as close as possible to the source.
Following the formalism described by Spruit (\cite{spruit})
and used by Sakurai (\cite{saku1}) we can deduce the relation between
$\omega$ and $\alpha$ that is relevant to fast rotators, which is
\begin{equation}
 \omega = \left(\frac{3}{2}\right)^{3/2} 
\left(1- \left(\frac{\alpha}{\alpha_*}\right)
^{2/3}\right)^{3/2}  
\label{OMspruit}
\end{equation}
This solution is represented in Fig.(\ref{figSPRUIT})
together with our numerical solutions on the equator.
When $\alpha$ decreases, the two curves get closer
while rotation parameter grows larger
and reaches $({3}/{2})^{3/2}$ when $\alpha = 0$.
The ultra-fast rotator limit is clear in this case.
Again we find that the larger the mass loss rate 
the smaller the rotation parameter. A disagreement arises with
the previous formula, since it does not allow 
$\alpha$ to be larger than unity. This shows the 
limited region of validity of this analytical 
approximation, which nevertheless describes quite well
the very fast rotator. 
Our analytical solutions tend to
Sakurai's results in the cold limit.
In our case the slow point is defined by
\begin{equation}
x_s = \left(\frac{g}{4\omega^2}\right)^{1/3} .
\end{equation}
This is equivalent to
\begin{equation}
\frac{\Omega^2 r_s^3}{G M} = 
\left(\frac{v_{pA}^2}{2}\right)^2 =
\tilde \omega_{Sakurai} 
\end{equation}
and the specific energy is
\begin{equation}
\epsilon = \left(\frac{3}{2}\right)
(2 \omega g)^{2/3} .
\end{equation}
Moreover, we can express those quantities
as functions of our input variables 
for very fast rotators and we get
\begin{equation}
r_s^3 = \frac{GM}{6} \left(\frac{A cos \theta^2}
{\omega \mu_0 \alpha}\right)^{2/3}
\end{equation}
and
\begin{equation}
\epsilon = \left(\frac{3}{2}\right)^{5/3}
\left(2 GM 
\left(\frac{\mu_0\Omega \alpha}{A \cos \theta^2} \right)^{1/3}
\right)^{2/3} .
\end{equation}
Thus we find the same results as those described by Spruit
in the cold limit, with more generality since we
solve the transversal force balance equation on the
Alfv\'en surface.

%%%%%%%%%%%%%%%%%%%%%%%%%%%%
%%%%%%%%%%%%%%%%%%%%%%%%%%%%
%%%%%%%%%%%%%%%%%%%%%%%%%%%%
\begin{figure}[htbp]
%\picplace{6cm}
\psfig{figure=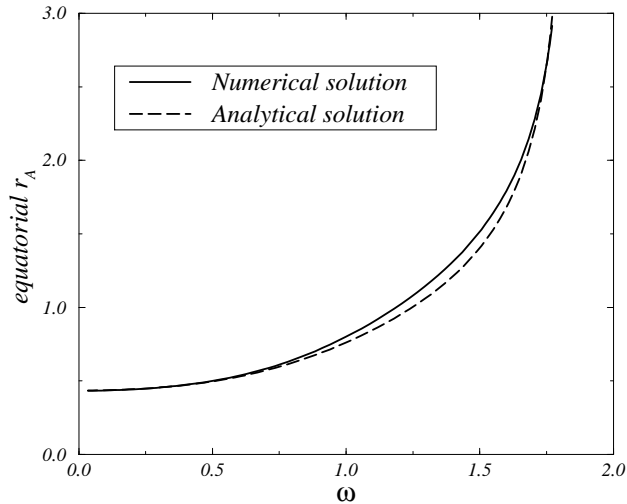,width=\linewidth}
\caption[ ]{
Comparisons of numerical results and analytical calculations
of the relation between the Alfv\`en radius on the equator and
$\omega$.
The upper curve (solid line) represents the numerical result while
the lower one  (dashed line) the analytical solution.
 The two curves are similar almost everywhere and particularly for
very slow and very fast rotators.
}
\label{figSPRUITra}
\end{figure}
%%%%%%%%%%%%%%%%%%%%%%%%%%%%

The shapes that we have found for the critical surfaces
differ from those obtained or assumed
in many other models. For example, Sakurai (\cite{saku2}) found
ellipsoidal shapes with his  numerical studies.  
Blandford \& Payne (\cite{bp})
have chosen conical Alfv\'enic surfaces in their 
self-similar model. Chan and Henriksen (\cite{chan}) 
and Sauty (\cite{sauty1})
used flat surfaces perpendicular to the axis of 
rotation. Our solutions show rather different shapes.
It appears to be difficult to 
capture realistic behaviors by modeling the critical
 surfaces with simple functions. 

Further developments within the framework of our model 
could be done to investigate the evolution of the angular momentum
of solar-mass stars during pre-main and main sequence phases.
Charbonneau (\cite{charbo}) recently derived a braking rate from a
Weber \& Davis (\cite{wd}) model of the solar-wind assuming a 
dynamo relationship where the stellar magnetic field scales
 linearly with the stellar angular velocity $\Omega_{\ast}$.
He finds that the braking rate, $d\Omega/dt$, asymptotically
scales as $\Omega_{\ast}^3$ at low $\Omega_{\ast}$, and
as $\Omega_{\ast}^2$ at high $\Omega_{\ast}$.
In the Weber \& Davis model,
the flatter $\Omega_{\ast}$-dependency of the 
braking rate for rapid rotators results from the change of the
 magnetic wind structure (Mestel (\cite{mestel}), Belcher \& MacGregor 
(\cite{belcher})).
That distinct braking laws apply to slow and fast rotators
is further supported by observations:
Skumanich's (\cite{skuma}) relationship ($\Omega_{\ast} \simeq t^{-1/2}$),
which follows from an $\Omega_{\ast}^3$ braking law, is
valid for slowly rotating dwarfs but fails for younger,
more rapidly rotating zero-age main sequence star
whose angular evolution is more appropriately described by an 
$\Omega_{\ast}^2$ braking law (Mermilliod \& Mayor (\cite{mermi})).
Thus our work takes place in a context of interesting
studies and some further investigations should
be done in this direction.

%%%%%%%%%%%%%%%%%%%%%%%%%%%%%%%%%%%%%%%%%%%%%%%%%%%%%%%%%%%%%
\subsection{A criterion for the classification of rotators}
%%%%%%%%%%%%%%%%%%%%%%%%%%%%%%%%%%%%%%%%%%%%%%%%%%%%%%%%%%%%%

With respect to the previous results,
it has been interesting to neglect the effects of the
thermal parameters and only to look at those of 
the rotation, of the magnetic field and of the mass loss rate.
It has been shown that an increase of $\Omega$ has effects 
similar to those of a decrease of $\alpha_0$, 
all the other parameters being fixed. 
This suggests that the ratio $\frac{\Omega}{\alpha_0}$
might be relevant for a study of magnetized rotators. 
That is the reason why we have plotted the relation between
the ratio ${\Omega}/{\alpha_0}$ and $\omega$.
First this gives a global relation between the input parameters
of the model that can be related to physical parameters 
given by observations (See appendix C)
and the variable $\omega$ that is a result of the calculation.

%%%%%%%%%%%%%%%%%%%%%%%%%%%%%%%%%%%%
\begin{figure}[htbp]
%\picplace{7cm}
\psfig{figure=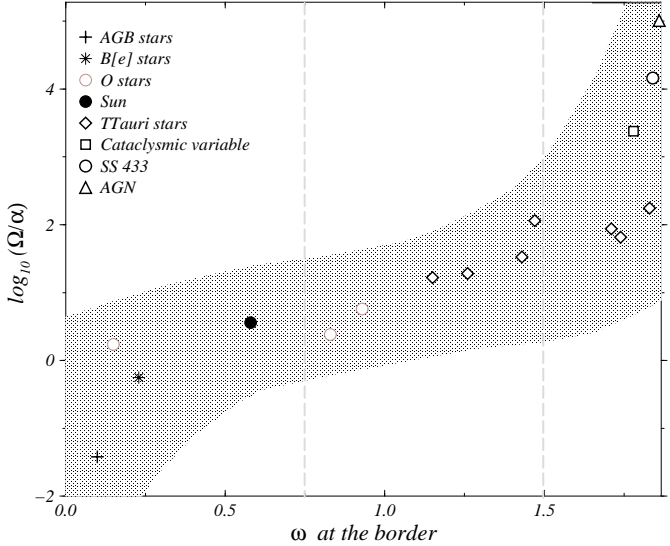,width=\linewidth}
\caption[]{Classification of the magnetic rotators.
the logarithm of the ratio $\frac{\Omega}{\alpha_0}$
 given by the observational literature is plotted against 
the rotation parameter $\omega$ obtained by the calculation
for several astrophysical objects.
 The grey zone represents the 
region of uncertainty over the input parameters.
The two vertical thin dashed line shows the 
approximate limits between slow, intermediate
and fast rotators. The limit for  ultra-fast rotators
is at $\omega = (\frac{3}{2})^{3/2}\approx 1.837$. 
}
\label{figCLASSIF}
\end{figure}
%%%%%%%%%%%%%%%%%%%%%%%%%%%%%%%%%%%%

We  find a continuity between slow and fast rotators.
Slow rotators correspond to $\omega \ll 1 $. 
Since this class contains the Solar wind, winds of
of B[e] stars, AGB stars, it can be identified with winds
in general.  In the intermediate regime that arbitrarily
corresponds to $0.75 < \omega < 1.5$ we find both
O stars and TTauri stars. This region clearly shows
a smooth transition between slow and fast rotators.
In the fast rotator zone
where $\omega$ is close to $\left(\frac{3}{2}\right)^{3/2}$
only object with jets, i.e. outflows of TTauri stars, 
of cataclysmic variable, of microquasars such as SS433 
and of active galactic nuclei, are to be found.
This shows two main types of objects, winds and jets
with a transition in between. 
One should notice an important difference between
the TTauri stars belonging to the intermediate zone
and those in the fast rotator regime.
They have larger mass loss rate ($> 10^{-7}  M_{\odot} yr^{-1}$)
 compared to the other ones (<$10^{-7}  M_{\odot} yr^{-1}$).
This could be due to a difference of ages for 
these TTauri stars. This suggests that $\dot M$
and $\omega$ are closely related.
Then our model could provide
a tool to see the evolution of this type of object
and to classify them.

From the input parameters  related
to astrophysical quantities given by observations
 and with the results of the model, 
we can classify outflows of magnetized rotators
into slow, intermediate and fast rotators.
We have also computed solutions for planetary
nebulae and Wolf-Rayet stars but, since the
origin of their wind is mainly radiative, 
they should not be compared with
the other objects. Nevertheless the numerical results 
obtained for  these types of astronomical
objects shows that they are definitely 
slow rotators following 
our classification.

\subsection{Limits of the model}

Our model possesses several advantages.
First it is not self-similar and secondly
it can accept any boundary conditions. Moreover
the simplifications allows us to find  fully analytical
solutions in limit cases and 
systems of equations which are easy to solve
numerically in all cases. The model is not simply
a continuous set of closely packed Weber-Davies solutions
 but it is made coherent with the Alfv\'en regularity condition
on the Alfv\'en surface, which retains part
of the cross-field balance. It also gives the
possibility to solve the Bernoulli equation
from the source to the fast point and finally
the asymptotic resolution of the outflow
is possible (See Lery et al. (\cite{lery2})) 
giving the possibility
to relate the properties of the source to the
asymptotic behavior of the outflow. 

The main  weaknesses however are the following.
First our model does not give
an exact solution to the system of coupled
Bernoulli and transfield equations. 
Secondly the geometry of 
the magnetic field lines is idealized.
This latter difficulty
could be avoided by iterating 
on the field geometry.

%%%%%%%%%%%%%%%%%%%%%%%%%%%%%%%%%%%%%%%%%%%%%%%%%%%%%%%%%%%%%
\section{Conclusions }
%%%%%%%%%%%%%%%%%%%%%%%%%%%%%%%%%%%%%%%%%%%%%%%%%%%%%%%%%%%%%

In this Paper, we have proposed a simplified set of the
 equations for  rotating magnetized outflows by assuming the shape of 
the poloidal magnetic field lines up to the fast magneto-sonic point.
Rather than solving the equilibrium perpendicular to the flux surfaces 
everywhere, solutions are found at the Alfv\`en point where it takes the 
form of the Alfv\`en regularity condition and
at the base of the flow. This constrains 
the transfield equilibrium in that the Alfv\`en
regularity condition is imposed and the regularity of
the magnetic surfaces at the Alfv\`en critical surface 
is ensured. In our model,
the outflow is parameterized by $\Omega(a)$, $Q(a)$ and $\alpha_0$
which is related to the mass loss rate.
We deduce three first integrals of the motion, the
specific angular momentum, the specific energy and the 
mass flux to magnetic flux ratio on a magnetic surface
from criticality and Alfv\`en regularity conditions.
Different profiles for the variation of 
the entropy and the rotation rate with 
respect to the magnetic flux have been considered.
The simplifications of the model allow to find
analytical behaviors  of the first integrals as 
well as the shape of the critical surfaces 
in limiting cases. We found good agreement between
analytical and numerical solutions.
For a given entropy, magnetic rotators can be characterized by 
the ratio of the rotation rate and 
the magnetic to mass flux ratio
on the polar axis, i.e. $\Omega/\alpha_0$.
This latter ratio is given by the boundary conditions.

Given the properties of the central emitting object,
the model  allows us to compute its corresponding 
dimensionless rotation parameter $\omega$.
Rotators can be defined as slow, intermediate or fast 
according to whether $\omega$ is much less or close to 
unity or near its maximum value for fast rotators, 
$(\frac{3}{2})^\frac{3}{2}$. 

Slow rotators have quasi-spherical critical surfaces
and a fast surface close to the Alfv\'en surface, their properties
strongly depend on the heating.

Fast rotators have distorted critical surfaces and
a fast surface far from the Alfv\'en surface  
(approaching infinity only in the zero temperature limit).
Their properties strongly depend on the magnetic flux and the rotation rate
yet always retain a slow rotator behavior near the axis of rotation.
They display a slow mode that acquires the character of a sound wave 
 as the rotation rate increases and are limited, in the present 
conical model, by $\omega = \frac{3}{2}^{\frac{3}{2}}$.

For all  types of rotators, it is found  using conical shapes 
that the mass loss rate has a lower limit for a given rotation rate.
The strongest effects on solutions are due to the rotation with 
respect the thermal and mass loss rate effects
for typical astrophysical values.
Given the angular velocity $\Omega(a)$ and the specific entropy $Q(a)$,
the solutions for the last first integrals of the motion, namely
the specific energy $E$, the specific angular momentum $L$ and the 
mass to magnetic flux ratio $\alpha$
 can be determined numerically and 
in some limiting cases analytically. 
Non-constant rotation cases allow two-velocities outflow types, with
possibly more complex solutions for different profiles of 
angular velocity and entropy.

This simplified model makes it possible to
 investigate the structure of outflows far from the 
magnetized rotator source, unconstrained by type of
boundary conditions, and without the need for 
self-similar assumptions.
How outflows extend into
the asymptotic region constitutes the subject
of the study of Paper II (Lery et al.(\cite{lery2}).
Paper III (Lery et al.(\cite{lery3}) 
will deal with a linear stability analysis of the
asymptotic equilibria.

%\begin{acknowledgements}
%\end{acknowledgements}

\appendix
\section{Cold rotators}

In the conical case it useful to
have the simplified equations 
for $\beta \ll 1$ since numerically
we have seen that this assumption
is quite acceptable as long as Q(a)
is not to big. These equations allow
easy analytical works. Particularly
for fast and slow rotators, one just have
to study the limit of variation of $\omega$
to get the equations for positions and densities
for the critical points.
Thus we have
\begin{equation}
y_s=\left({\frac
{\beta \left(\gamma-1\right) x_s^4}
{2\left( 1+\omega^2 x_s^2
\left(1-x_s^2\right)^2\right)}}
\right)^{\frac{-1}{\gamma+1}} 
\label{si1}\end{equation}
\begin{equation}
y_f ={\frac
{1}
{1+\left(\omega x_f\left(x_f^2-1\right)\right)^{\frac{2}{3}}}}
\label{si2}\end{equation}
One can deduce a reduced form for the relative energy
and the relative gravity that only depend on 
$x_s$, $x_f $
and on two parameters $\omega, \beta$ 
\begin{eqnarray}
\epsilon & = &
\beta \left( 1 + \omega^{\frac{2}{3}} x_f^2
( 1-x_f^{-2} )^
{\frac{2}{3}}\right)^{1-\gamma} \nonumber \\
& & + \frac 
{6 \omega^{\frac{4}{3}}}
{\left( 1-x_f^{-2} \right)^{\frac{2}{3}}}
-3\omega^{\frac{4}{3}}\left(1-x_f^{-2}
 \right)^{\frac{4}{3}}
 \nonumber \\
& & + x_f^{-2}
\{ -\omega^2 
-6 \omega^{\frac{2}{3}}\left(1-x_f^{-2} 
\right)^{\frac{2}{3}}
\nonumber\\
& &
+\frac
{3 \omega^{\frac{2}{3}}}
{\left(1-x_f^{-2} \right)^{\frac{4}{3}}}
-\frac
{4\omega^{\frac{4}{3}}}
{\left(1-x_f^{-2} \right)^{\frac{2}{3}}}
\}
 \nonumber\\
& & + x_f^{-4}
\{-3-\frac{2\omega^{\frac{2}{3}}}{\left(1-x_f^{-2} 
\right)^{\frac{4}{3}}}
-\frac{2\omega^{\frac{4}{3}}}{\left(1-x_f^{-2} 
\right)^{\frac{2}{3}}}
\} \nonumber \\
 & & + x_f^{-6}
\{ \frac{-\omega^{\frac{2}{3}}}{\left(1-x_f^{-2} 
\right)^{\frac{4}{3}}}
\} 
\end{eqnarray}

\begin{eqnarray}
\epsilon &=&
\omega^2 (2-3 x_s^2) \nonumber\\
&&  +
\left(
\frac
{\beta\left(\gamma-1\right)x_s^4}
{2\left(1+\omega^2 x_s
^2(1-x_s^2)^2\right)}\right)
^{\frac{2}{\gamma+1}}  \nonumber\\
&& 
\left(\frac
{5-3\gamma}{(\gamma-1) x_s^4}\right) \nonumber\\
&&  +
\left(
\frac
{\beta\left(\gamma-1\right)x_s^4}
{2\left(1+\omega^2 x_s
^2(1-x_s^2)^2\right)}\right)
^{\frac{2}{\gamma+1}}   \nonumber\\
&&
\left(\frac
{\omega^2 x_s^2 
(1-x_s^2) 
\{ 2(1-x_s^2)-(\gamma-1) 
(3 x_s^2+1) \} }
{(\gamma-1) x_s^4}\right) 
\end{eqnarray}

\begin{eqnarray}
{\frac{g}{2}} & = & \frac
{2\omega^{\frac{4}{3}}}
{({\frac {r_{A}}{r_{f}}})}
\left(1-\frac
{1}
{\left(1-({\frac {r_{A}}{r_{f}}})^2 \right)^{\frac{2}{3}}}\right) \nonumber \\
& & + ({\frac {r_{A}}{r_{f}}}) 
\left( \omega^2 
-\frac
{\omega^{\frac{2}{3}}}{\left(1-({\frac {r_{A}}{r_{f}}})^2 \right)^{\frac{4}{3}}}
+\frac{4\omega^{\frac{2}{3}}}
{\left(1-({\frac {r_{A}}{r_{f}}})^2 \right)^{\frac{2}{3}}} 
\right) \nonumber \\
& & + ({\frac {r_{A}}{r_{f}}})^3
\left( \frac{2}{\left(1-({\frac {r_{A}}{r_{f}}})^2 \right)^{\frac{4}{3}}}
+ \frac
{2\omega^{\frac{4}{3}}}{\left(1-({\frac {r_{A}}{r_{f}}})^2 \right)^{\frac{2}{3}}}
\right) \nonumber \\
& & + ({\frac {r_{A}}{r_{f}}})^5
\left(\frac
{\omega^{\frac{2}{3}}}{\left(1-({\frac {r_{A}}{r_{f}}})^2 \right)^{\frac{4}{3}}}
\right)
\end{eqnarray}

\begin{eqnarray}
\frac{g}{2} &=&\omega^2 x_s^3+
\left(
\frac{\beta\left(\gamma-1\right)x_s^4}
{2\left(1+\omega^2 x_s^2
(1-x_s^2)^2\right)}\right)
^{(\frac{2}{\gamma+1})}\nonumber\\
&&\left(
\frac{2+\omega^2 x_s^2
(1-x_s^2)(1+x_s^2)}
{x_s^3}\right)
\label{simpl3}
\end{eqnarray}

\section{Profiles of $\Omega(a)$ and $Q(a)$}
Definitions of parameters:
\begin{itemize}
\item[$\bullet$]$\Omega_0$: angular velocity
\item[$\bullet$]$a$: magnetic flux
\item[$\bullet$]$N$: large integer
\item[$\bullet$]$B$ and $C$: constants 
\item[$\bullet$]$C_1$ and $C_2$: constants with $C_1 +C_2 \leq 1$
\end{itemize}
\paragraph{Differential rotators}

\begin{itemize}
	\item Rigid body rotation
\begin{equation}
	\Omega = \Omega_0
\label{omRIG}
\end{equation}
	\item Keplerian rotation
\begin{eqnarray}
	\Omega &=& \Omega_0 
(\frac
{({1-a}{\sqrt{a (2 -a)}})^{3/2}}
{1+\exp(-N a + N/10) } \nonumber \\
& & +
\frac{1}{(1+\exp(-N a + N/10)) }
)
\label{omKEP}
\end{eqnarray}

	\item Stellar type rotation
\begin{equation}
	\Omega = \Omega_0 (1 -B \sin^2\theta -C \sin^4\theta)
\label{omSOL}
\end{equation}
with
\begin{equation}
               \theta =  \arccos\left(\sqrt{a(2-a)}\right)
\end{equation}

	\item Decreasing stellar type rotation
\begin{equation}
	\Omega = \Omega_0 (1 -B \cos^2\theta -C \cos^4\theta)
\label{omSOLneg}
\end{equation}

	\item Two velocities rotation ($\Omega_0$-0)
\begin{equation}
	\Omega = \frac{\Omega_0}{\exp(N(a-\frac{1}{2}))+1-\exp(-\frac{N}{2})} 
\label{om10}
\end{equation}

	\item Two velocities rotation ($\Omega_0$-$\Omega_1$)
\begin{equation}
	\Omega = \Omega_0 \left(C_1+\frac{C_2}
	{\exp(N(a-\frac{1}{2} ))+1-\exp(-\frac{N}{2})}\right)
\label{om1C1}
\end{equation}

\end{itemize}

\paragraph{Variations of $Q(a)$}

\begin{itemize}

	\item Constant entropy
\begin{equation}
	Q = Q_0
\label{omQcon}
\end{equation}

	\item Two entropies  $ (Q_0-\frac{Q_0}{2})$
\begin{equation}
	Q = Q_0 \left(0.5+\frac{0.5}
	{\exp(N(a-\frac{1}{2} ))+1-\exp(-\frac{N}{2} )}\right)
\label{omQ05}
\end{equation}

	\item Increasing entropy
\begin{equation}
	Q = Q_0 (1 -B \sin^2\theta -C \sin^4\theta)
\label{omQneg}
\end{equation}
with
\begin{equation}
               \theta =  \arccos\left(\sqrt{ (2-a)}\right)
\end{equation}

	\item Decreasing entropy
\begin{equation}
	\Omega = \Omega_0 (1 -B \cos^2\theta -C \cos^4\theta)
\label{omQpos}
\end{equation}

\end{itemize}

\section{Examples of input parameters for astrophysical objects}

\begin{table*}
\caption{Examples of input parameters for astrophysical objects}
%\picplace{15cm}
\begin{tabular}{lrllllllllll}
\hline
\hline
\noalign{\smallskip}
Object type & $log{(\frac{\Omega}{\alpha})}$&$\omega$& 
${\overline \Omega}$ & ${\overline\alpha}$ & ${\overline Q}$ 
& $\dot M$ & $M$ & $R$ & $T$ & $n$ & $B$ \\
 & & & & & & $(M_{\odot} yr^{-1})$  &  
$(M_{\odot}) $  &  $(R_{\odot})$& (K) & $( cm^{-3})$ & (G)  \\
\noalign{\smallskip}
\hline
\noalign{\smallskip}
AGB   $^{\rm a}$ & -1.70  & 0.07 &0.001& 0.05  & 0.26& $2\times10^{-6}$  & 5  & 500 & 2800           & $10^{10}$& 0.1            \\
B[e]  $^{\rm b}$ & -0.53  & 0.20 & 0.5 & 1.7   & 0.2 & $2\times10^{-6}$  & 37 & 86  & $2\times10^4$  & $10^{10}$& 1              \\
O5 V  $^{\rm c}$ & -0.05  & 0.12 & 7.0 & 7.85  & 20.4& $2\times10^{-10}$ & 1  & 1   & $2\times10^6$  & $10^8$   & 1              \\
O3 III$^{\rm c}$ &  0.10  & 0.80 & 0.56& 0.44  & 0.8 & $6.2\times10^{-6}$& 50 & 19.7& $6.5\times10^4$& $10^{10}$& $2\times10^2$  \\
O3 III$^{\rm c}$ &  0.48  & 0.91 & 0.56& 0.18  & 0.8 & $6.2\times10^{-6}$& 50 & 19.7& $6.5\times10^4$& $10^{10}$& $1.6\times10^3$\\
Sun   $^{\rm d}$ &  0.26  & 0.55 & 3.1 & 1.7   & 2.1 & $10^{-14}$        & 1  & 1   & $10^6$         & $10^{10}$& $10^3$         \\
DF Tau$^{\rm e,h}$& 0.94  & 1.12 & 0.8 & 0.09  & 0.07& $1.3\times10^{-7}$& 2  & 2.5 & $4\times10^{3}$& $10^4$   & $2\times10^3$  \\
GG Tau$^{\rm e}$ &  1.00  & 1.23 & 1.0 & 0.1   & 0.03& $4\times10^{-7}$  & 0.8& 3.5 & $10^4$         & $10^{8}$ & $2\times10^3$  \\
BP Tau$^{\rm e}$ &  1.24  & 1.41 & 1.8 & 0.1   & 0.05& $2\times10^{-7}$  & 0.8& 3.0 & $9\times10^3$  & $10^4$   & $1\times10^3$  \\
RY Tau$^{\rm e}$ &  1.53  & 1.71 & 2.2 & 0.065 & 0.02& $7.5\times10^{-8}$& 2  & 2.7 & $8\times10^3$  & $10^4$   & $1\times10^3$  \\
DS Tau$^{\rm e}$ &  1.66  & 1.68 & 2.9 & 0.022 & 0.05& $6.5\times10^{-8}$& 1  & 1.8 & $9\times10^3$  & $10^6$   & $2\times10^3 $ \\
T Tau $^{\rm e}$ &  1.78  & 1.44 & 1.23& 0.02  & 0.07& $1.1\times10^{-7}$& 2  & 4   & $7\times10^3$  & $10^{10}$& $2\times10^3$  \\
SU Aur$^{\rm e}$ &  1.96  & 1.80 & 50  & 0.05  & 3.0 & $2\times10^{-8}$  & 2.25& 3.6& $2\times10^{5}$& $10^8$   & 300            \\
C.V.  $^{\rm f}$ &  3.10  & 1.75 & 11  & 0.008 & 1.4 & $2\times10^{-10}$ & 1  & 1.2 & $10^6$         & $7\times10^{13}$ &$10^6$  \\
SS 433$^{\rm g}$ &  3.88  & 1.81 & 65  & 0.009 & 0.1 & $10^{-4}$         & 10 & 0.2 & $10^5$         & $10^{10}$& $10^9 $        \\
A.G.N.$^{\rm g}$ &  4.73  & 1.83 & 160 & 0.003 & 0.7 & $10^{-4}$         &$10^9$ &150 &$2\times10^5$ & $2\times10^6$& $10^4$     \\
\noalign{\smallskip}
\hline
$^{\rm a}$ Livio (\cite{livio}).  \\
$^{\rm b}$ Cassinelli et al. (\cite{cassi}). \\
$^{\rm c}$ Mc Gregor (\cite{mcgregor}).\\
$^{\rm d}$ Priest (\cite{priest}).  \\
$^{\rm e}$ Bertout et al. (\cite{bertout}).\\
$^{\rm f}$ Murray \& Chiang (\cite{murray}). \\
$^{\rm g}$ Bremer (\cite{bremer}).  \\
$^{\rm h}$ Thi\'ebaut et al. (\cite{thieb}).
\end{tabular}
\end{table*}


\begin{thebibliography}{}

\bibitem[1976]{belcher}
Belcher, J.W., MacGregor, K.B., 1976, ApJ 210, 498.

\bibitem[1988]{bertout}
Bertout, C., et al., 1988, ApJ, 330, 350.

\bibitem[1982]{bp}
Blandford, R.D., Payne, D.G., 1982, MNRAS, 199, 883.

\bibitem[1996]{bremer}
Bremer, M., 1996, cghr.conf. B., 
Astrophysics and space science library,
ed. Bremer, M. (Dordrecht: Kluwer).

\bibitem[1994]{cao}
Cao, X., Spruit, H.C., 1994, A\&A, 287, 80.

\bibitem[1989]{cassi}
Cassinelli, J.P., et al., 1989,
in {\it Physics of luminous blue variables}, ed. K. Davidson et al.
(Dordrecht:Kluwer), 121. 

\bibitem[1980]{chan}
Chan, K.L., Henriksen, R.N., 1980, ApJ, 241, 534.

\bibitem[1992]{charbo}
Charbonneau, P., 1992, in {\it Seventh Cambridge Workshop on Cool stars,
Stellar systems and the Sun}, ASP Conf. Ser., ed. M.S. Giampapa \&
J.A. Bookbinder, 26, 417.

\bibitem[1994]{conto}
Contopoulos, J., Lovelace, R.V.E., 1994, ApJ, 429, 139.

\bibitem[1993]{ferr1}
Ferreira, J., Pelletier, G., 1993a, A\&A, 276, 625.

\bibitem[1993]{ferr2}
Ferreira, J., Pelletier, G., 1993b, A\&A, 276, 637.

\bibitem[1997]{ferr3}
Ferreira, J., 1997, A\&A, 319, 340.

\bibitem[1996]{fiege}
Fiege, J.D., Henriksen, R.N., 1996, MNRAS, 281, 1038.

\bibitem[1982]{hart}
Hartmann, L., Mc Gregor, K.B., 1982, ApJ, 259, 180.


\bibitem[1994]{henrik}
Henriksen, R.N., Valls-Gabaud, D., 1994, MNRAS, 266, 681.

\bibitem[1989]{HN}
Heyvaerts, J., Norman, C., 1989, ApJ, 347, 1055.

\bibitem[1989]{kennel}
Kennel, C.F., Fujimura, F.S., Okamoto, I., 
Geophys. Astrophys. Fluid Dynamics, 1983, 26, 147.

\bibitem[1989]{ko}
K\"onigl, A., 1989, ApJ , 342, 208.

\bibitem[1989]{koupe}
Koupelis, T., Van Horn, H.M., 1989, ApJ , 342, 146.

\bibitem[1990]{koupe2}
Koupelis, T., 1990, ApJ , 363, 79.

\bibitem[1998b]{lery2}
Lery, T., Heyvaerts, J., Appl, S., Norman, C.A., 
1998b, A\&A, in preparation (Paper II).

\bibitem[1998c]{lery3}
Lery, T., Appl, S., Heyvaerts, J.,
1998c, A\&A, in preparation (Paper III).

\bibitem[1995]{li}
Li, Z. Y., 1995, ApJ, 444, 848.

\bibitem[1994]{livio}
Livio, M., 1994 in {\it Circumstellar Media in late stages of stellar
evolution}, ed. R. Clegg et al. (Cambridge Univ. Press), 35.

\bibitem[1996]{mcgregor}
Mc Gregor, K.B., 1996, in K.C. Tsinganos (ed.), 
{\it Solar \& Astrophysical Magnetohydrodynamic Flows}, Kluwer, 301.

\bibitem[1990]{margu}
Margulis, M., et al., 
1990, ApJ, 352, 615.

\bibitem[1990]{mermi}
Mermilliod, J.C., Mayor, M., 1990, A\&A, 237, 61.

\bibitem[1968]{mestel}
Mestel, L., 1968, MNRAS, 138, 359.

\bibitem[1987]{ms}
Mestel, L., Spruit, H.C., 1987, MNRAS, 226, 57.


\bibitem[1969]{michel}
Michel, F.C., 1969, ApJ, 158,727.

\bibitem[1996]{murray}
Murray, N., Chiang, J., 1996, in {\it Nature}, 382, 789. 

\bibitem[1994]{naji}
Najita, J. R., Shu, F.H., 1994, ApJ, 429, 808.

\bibitem[1978]{oka}
Okamoto, I., 1978, MNRAS, 185, 69.

\bibitem[1995]{ouyed}
Ouyed, R., 1995, BAAS, 27, 1317.

\bibitem[1997]{ouyedp}
Ouyed, R., Pudritz, R.,1997, ApJ, 482, 712.

\bibitem[1963]{parker}
Parker, E.N., 1963, {\it Interplanetary dynamical processes},
 New York, Interscience Publishers.

\bibitem[1992]{pellpud}
Pelletier, G., Pudritz, R.E., 1992, ApJ, 394, 117.

\bibitem[1987]{priest}
Priest, E.R., 1987, {\it Solar magneto-hydrodynamics},
D. Reidel publishing Company (Dordrecht).

\bibitem[1986]{PN}
Pudritz, R. E., Norman, C. A., 1986,  ApJ, 301, 571.

\bibitem[1992]{RHP}
Richer, J.S., Hills, R.E., Padman, R., 
1992, MNRAS, 254, 525.

\bibitem[1994]{rosso}
Rosso, F., Pelletier, G., 1994, A\&A, 287, 325.

\bibitem[1985]{saku1}
Sakurai, T., 1985, A\&A, 152, 121.

\bibitem[1987]{saku2}
Sakurai, T., 1987, PASJ, 39, 821.

\bibitem[1993]{sauty1}
Sauty, C. 1993 PhD Thesis, University of Paris VII.

\bibitem[1994]{sautyt}
Sauty, C., Tsinganos, K., 1994, A\&A, 287, 893.

\bibitem[1988]{shul}
Shu, F.H., Lizano, S. Ruden, S.P. and Najita, J. 1988, ApJ, 328, L19

\bibitem[1994]{shu94}
Shu, F.H., Najita, J., Ostriker, E., Wilkin, F., Ruden, S. 
and Lizano, S., 1994, ApJ, 429, 781.


\bibitem[1972]{skuma}
Skumanich, A., 1972, ApJ, 171, 565.

\bibitem[1994]{spruit}
Spruit, H.C., 1994, {\it Cosmical Magnetism}, NATO ASI Series C.
eds. D. Lynden-Bell (Kluwer), 422, 33.

\bibitem[1973]{suess}
Suess, S.T., Nerney, S.F., 1973, ApJ, 184, L17.

\bibitem[1995]{thieb}
Thi\'ebaut, E., Balega, Y., Balega, I., Belkine, I.,
Bouvier, J., Foy, R., Blazit, A. and Bonneau, D., 
1995, A\&A, 304, 17.

\bibitem[1992]{tsin1}
Tsinganos, K., Sauty, C., 1992, A\&A, 255, 405.

\bibitem[1991]{tsin2}
Tsinganos, K., Trussoni, E., 1991, A\&A, 249, 156.

\bibitem[1985]{uchida}
Uchida, Y., Shibata, K., 1985, PASJ, 37, 515.

\bibitem[1993]{wk}
Wardle, M., K\"onigl, A., 1993, ApJ, 410, 218.

\bibitem[1967]{wd}
Weber, E.J., Davis, L., 1967, ApJ, 148, 217.


\end{thebibliography}
\end{document}